\documentclass[twocolumn,superscriptaddress,showpacs,prc,
twoside,floatfix]{revtex4-1}
\usepackage{dcolumn}
\usepackage{graphicx,subfigure}
\usepackage{xcolor}
\usepackage{soul}

\usepackage{longtable}
\graphicspath{{calculate_potential/}}
\newcommand {\la} {\langle}
\newcommand {\ra} {\rangle}
\newcommand {\beq} {\begin{eqnarray}}
\newcommand {\eeqn} [1] {\label{#1} \end{eqnarray}}%
\newcommand {\eol} {\nonumber \\}
\newcommand {\ve} [1] {\mbox{\boldmath $#1$}}

\newcolumntype{P}[1]{>{\centering\arraybackslash}p{#1}}

\begin{document}
%
%

\title{Implications for (d,p) reaction theory from nonlocal dispersive optical model analysis of $^{40}$Ca(d,p)$^{41}$Ca.}
\date{\today}
\author{S. J. Waldecker
}                 

\affiliation{
Department of Chemistry and Physics, University of Tennessee at Chattanooga, 
615 McCallie Avenue, Chattanooga, TN 37403
}

\author{N. K. Timofeyuk}

\affiliation{
 Department of Physics, Faculty of Engineering and Physical Sciences, 
University of Surrey, Guildford,
Surrey GU2 7XH,  UK
}

\begin{abstract}
The nonlocal dispersive optical model (NLDOM) nucleon potentials are used for the first time in the adiabatic analysis of a (d,p) reaction to generate distorted waves both in the entrance and exit channels. These potentials were designed and fitted by Mahzoon \emph{et al.} [Phys. Rev. Lett. 112, 162502 (2014)] to constrain relevant single-particle physics in a consistent way by imposing the fundamental properties, such as nonlocality, energy-dependence and dispersive relations, that follow from the complex nature of nuclei. However, the NLDOM prediction for the $^{40}$Ca(d,p)$^{41}$Ca cross sections at low energy, typical for some modern radioactive beam ISOL facilities, is about 70\% higher than the experimental data despite being reduced by the NLDOM spectroscopic factor of 0.73. This overestimation comes most likely either from insufficient absorption or due to constructive interference between ingoing and outgoing waves. This indicates strongly that additional physics arising from many-body effects is missing in the widely used current versions of (d,p) reaction theories.

\end{abstract}
\maketitle

\date{\today}

\section{Introduction}

One nucleon transfer reactions have been a tool for nuclear spectroscopic studies for half of a century. Today, they are used  in experiments with radioactive beams and among them the (d,p) reactions are perhaps the most popular choice. Analysis of these reactions relies on (d,p)  reaction theory, which is traditionally  
either  the distorted-wave Born approximation (DWBA) \cite{Satchler} or   adiabatic distorted wave approximation (ADWA)  \cite{JT}, the latter being a computationally inexpensive way of taking deuteron breakup into account.
The deuteron breakup means that the (d,p) amplitude should contain the $A+n+p$ degrees of freedom explicitly, which requires solving the $A+n+p$ three-body Schr\"odinger equation. Several methods exist to solve this equation exactly, the CDCC \cite{Mor09, Del15} and the Faddeev approach \cite{Del09a} being the most used. The usual assumption in these calculations is that the $A+n+p$ Hamiltonian contains the $p-A$ and $n-A$ potentials (often taken at half the deuteron incident energy) that describe nucleon elastic scattering. However,
it has been shown in \cite{Joh14} that the $n-A$ and $p-A$ potentials for the $A+n+p$ problem are very complicated objects which depend on the position and the energy of the third nucleon and are not equal to optical potentials taken at half the deuteron energy. It was also shown in \cite{Joh14} that in the case of (d,p) reactions the averaging over the first Weinberg component  (which is the same as making the adiabatic approximation) results in a simple prescription for choosing the $n-A$ and $p-A$ potentials appropriate for analysis of (d,p) reactions. This prescription is possible due to the main contribution to the (d,p) amplitude coming from small $n-p$ separations.

The prescription in \cite{Joh14} says that within the Feshbach formalism, the $n-A$ and $p-A$ potentials should be nonlocal energy-dependent potentials evaluated at half the deuteron incident energy plus half of
the $n-p$ kinetic energy in deuteron averaged over the $n-p$ potential, which is about 57 MeV. After evaluation of these potentials, they should be treated as energy-independent and nonlocal.  A simple recipe to include such potentials into the available (d,p) reaction scheme, based on the local energy approximation, is given in \cite{Tim13b}. 

At the time when \cite{Joh14} was written, only one energy-dependent nonlocal potential had been known \cite{GRZ}, derived from Watson multiple scattering theory. It has an energy-independent real part and an energy-dependent imaginary potential. Soon after the publication of \cite{Joh14}, a nonlocal version of the dispersive optical model (NLDOM) became available for $^{40}$Ca \cite{NLDOM}. The nonlocal structure of NLDOM is more complicated than that from previous nonlocal optical potentials in that it is described by seven different nonlocality parameters. Based on the nucleon self-energy from Green's function many-body theory, the NLDOM potential contains both real and imaginary dynamic terms that are connected through a dispersion relation, which enforces causality and links the negative and positive energy regions. This dispersion relation is important for constraining the NLDOM parameters with both scattering and bound-state data while simultaneously providing a good description of these data. The potential from Ref.~\cite{GRZ} was also constrained with both scattering and bound-state data but without incorporating the dispersion relation. 

In this paper, we analyze the $^{40}$Ca(d,p)$^{41}$Ca reaction at 11.8 MeV using NLDOM to generate the distorting potentials in both the entrance and exit channels. The NLDOM potential has already been used in \cite{Ross15} to calculate the $^{40}$Ca(p,d)$^{39}$Ca cross sections but only within the DWBA (which means neglecting deuteron breakup) and no comparison to the experimental data was made. Our choice of the reaction is due to the availability of the  p-$^{40}$Ca and n-$^{40}$Ca optical potentials needed to construct the d-$^{40}$Ca potential. The choice of the deuteron energy is due to several radioactive beams facilities existing in the world that use this low-energy range. Also, it is this low-energy range where the dispersive relations cause the most prominent effects in the energy behavior of the optical potential. In addition, at these energies, spin-orbit effects and finite range effects can be neglected and the prescription from \cite{Joh14} should be valid.

In Sec.~\ref{sec:nldom}, we review the NLDOM and show that, similar to the standard Perey-Buck case, a local-equivalent potential exists for NLDOM and a generalization of the Perey factor can be introduced. In a similar fashion, we show in Sec.~\ref{sec:adwa} that
 the d-$^{40}$Ca distorting potential can be constructed by extending the local scheme proposed in \cite{Tim13b} to the case with several nonlocality parameters. We summarize the adiabatic approximation in lowest order and introduce first order corrections. In Sec.~\ref{sec:results} we calculate the cross section of the  $^{40}$Ca(d,p)$^{41}$Ca reaction at 11.8 MeV and show that, using the prescription from ~\cite{Joh14}, the NLDOM strongly overestimates the experimental data. In Sec.~\ref{sec:con}, we discuss the implications for the (d,p) reaction theory following from our analysis. 

\section{Nonlocal DOM potential and nucleon scattering}
\label{sec:nldom}
The NLDOM potential from Ref.~\cite{NLDOM} models the irreducible nucleon self-energy $\Sigma(\ve{r}, \ve{r}'; E)$ with real and imaginary parts that are both explicitly nonlocal. The potential contains eight terms, which were constrained with both bound-state and scattering data associated with $^{40}$Ca. It is written in the form
\beq
\Sigma(\ve{r}, \ve{r}'; E) &= &U_{HF}^{vol 1}( \tilde{r}) H(\ve{x}; \beta_{vol_1}) \nonumber \\
& + &U_{HF}^{vol 2}( \tilde{r} ) H(\ve{x}; \beta_{vol_2}) \nonumber \\
& + &U_{HF}^{wb}( \tilde{r} ) H(\ve{x}; \beta_{wb}) \nonumber \\
& + &U_{dy}^{sur +}( \tilde{r}; E) H(\ve{x}; \beta_{sur +}) \nonumber \\
& + &U_{dy}^{sur -}(\tilde{r}; E) H(\ve{x}; \beta_{sur - }) \nonumber \\
& + &U_{dy}^{vol +}(\tilde{r}; E) H(\ve{x}; \beta_{vol +}) \nonumber \\
& + &U_{dy}^{vol -}(\tilde{r}; E) H(\ve{x}; \beta_{vol -}) \nonumber \\
& + &U_{so}(r; E) 
\eeqn{DOMpot}
where $\tilde{r} = |\ve{r} + \ve{r}'| / 2$ and $\ve{x} = \ve{r}' - \ve{r}$. Following Perey and Buck \cite{PB}, the nonlocality function $H$ is assumed to be of the form
\beq
H(\ve{x};\beta) = \exp(-\ve{x}^2/\beta^2) / (\pi^{3/2}\beta^3),
\eeqn{gauss}
where $\beta$ is the nonlocality range.

In Eq. (\ref{DOMpot}) the $U_{HF}$ terms represent the static part of the self-energy and are purely real. For this reason, these terms are referred to as parts of the Hartree-Fock (HF) potential, but technically they do not form the true HF potential, because in practice a subtracted dispersion relationship is used \cite{NLDOM}. The $U_{dy}$ terms represent the dynamic part of the self-energy and are complex. The real part of the dynamic self-energy is determined completely from the dispersion integral of the imaginary part. The dynamic self-energy consists of surface and volume terms, and these have different nonlocalities for energies above the Fermi energy $E_F$ (denoted with a '+' sign) and energies below $E_F$ (denoted by a '-' sign).  The inclusion of several nonlocality parameters was based on the microscopic calculation in Ref.~\cite{Dussan11}, which indicated different degrees of nonlocality in different energy regions. Table~\ref{Tbl:DOM_betas} shows the value of each nonlocality parameter. Note that some of these parameters are about twice as large as those known from traditional Perey-Buck potentials.  The $U_{so}$ term is the spin orbit potential, which was assumed to be local. It has a weak energy dependence that only becomes important at high energies.  

Overlap functions can also be generated with the dispersive optical model. For discrete states in the $A+1$ and $A-1$ systems, one can show that these overlap functions obey a Schr{\"o}dinger-like equation~\cite{Dickhoff08} with the nucleon self-energy taking the role of an effective potential. In order to use the NLDOM overlap function in the analysis of $^{40}$Ca (d,p)$^{41}$Ca reactions, the calculated binding energy of the $0f_{7/2}$ neutron level in $^{40}$Ca must match the experimental one of $8.36$ MeV. However, in \cite{NLDOM}, such a constraint was not employed. For the purposes of this study, some of the parameters were refit in order to reproduce the experimental binding energy of the $0f_{7/2}$ neutron level. Only the parameters associated with the $U_{HF}$ terms were refit, and they were constrained by elastic scattering, charge density, and energy level data. All other parameters were unchanged. The new parameters are shown in Table~\ref{Tbl:DOM_HF} and compared with those from the analysis in \cite{NLDOM}. The quality of the new fit is comparable to that obtained in \cite{NLDOM}.

\begin{table}
\caption{ NLDOM nonlocality parameters.}
\label{Tbl:DOM_betas}

\begin{tabular}{P{3.0 cm}P{3.0 cm}}
\hline
\hline
Parameter & Nonlocality (fm) \\ \hline 
$\beta_{vol_1}$ & 0.84 \\
$\beta_{vol_2}$ & 1.55 \\
$\beta_{wb}$ & 1.04 \\
$\beta_{sur +}$ & 0.94 \\
$\beta_{sur -}$ & 2.07 \\
$\beta_{vol +}$ & 0.64 \\
$\beta_{vol -}$ & 0.81 \\
\hline
\hline
\end{tabular}
\end{table}

\begin{table}
\caption{Adjusted HF parameters used in the present fit. For a description of these parameters, refer to \cite{NLDOM}.} 
\label{Tbl:DOM_HF}

\begin{tabular}{P{2.5 cm}P{2.5 cm}P{2.5 cm}}
\hline
\hline
Parameter & New value & Old value \\ \hline 
$V_{HF}^0$ [MeV] & 106.15 & 100.06 \\
$r_{HF}$ [fm] & 1.14 & 1.10\\
$a_{HF}$ [fm] & 0.58 & 0.68\\
$\beta_{vol_1}$ [fm]& 0.84 & 0.66 \\
$\beta_{vol_2}$ [fm] & 1.55 & 1.56 \\
$x_1$ & 0.48 & 0.48 \\
$V^0_{wb}$ [MeV] & 12.5 & 15.0\\
$\rho_{wb}$ [fm] & 2.06 & 1.57 \\
$\beta_{wb}$ [fm] & 1.04 & 1.10 \\ 
\hline
\hline
\end{tabular}
\end{table}

The nonlocal terms in Eq.~(\ref{DOMpot}) can be written more succinctly as
\beq
\Sigma_N(\ve{r}, \ve{r}'; E) = \sum_i {U_{NA, i}(\tilde{r}) H_i(x)},
\eeqn{DOMpot2}
where the energy dependence of the dynamic terms is implied, and $N = p, n$ for proton and neutron potentials, respectively.

In this paper we will construct an effective local model for the deuteron-target adiabatic potential using NLDOM. Therefore, we first evaluate the effective local potential model for nucleon scattering. Following the procedure in \cite{PB} for transforming a nonlocal potential to a local equivalent, one finds that for a nonlocal potential with multiple nonlocalities of the Perey-Buck type the local-equivalent potential can be found by solving the transcendental equation
\beq
U_{loc}(r) = \sum_{i}U_{NA, i}(r)\exp\left[-\frac{\mu_N\beta_i^2}{2\hbar^2}(E - U_{loc}(r))\right]
\eeqn{uloc_multi}
where $\mu_N$ is the reduced mass of the $N+A$ system. This equation is obtained in the lowest order of the expansion  of $\Sigma_N(\ve{r}, \ve{r}'; E)$ over $\ve{x}$ and corrections to any order can be built systematically using developments from \cite{Fid}. In particular, the first order correction is
\beq
\Delta U_N = \frac{\hbar^2}{\mu_N} \left[ \left(\frac{\nabla f}{f}\right)^2-\frac{1}{2} \frac{\nabla^2 f}{f}\right],
\eeqn{Delta_U}
where the function $f(r)$ is the Perey factor explained below. For proton scattering,  equation (\ref{uloc_multi}) must be corrected by reducing  the centre-of-mass energy $E$ in the r.h.s.
of Eq. (\ref{uloc_multi}) by the local Coulomb interaction $V_{coul}(r)$.

According to \cite{Fid}, the local spin-orbit term must also be corrected when transforming to a local-equivalent potential. For the present case of a potential with multiple nonlocality parameters, the new spin-orbit term, $U^{le}_{so}$, of the local-equivalent potential is 
\beq
U_{so}^{le} = U_{so} / (1 - U_1)
\eeqn{leso}
where 
\beq
U_1 = \left(1 - \sum_{i}\frac{\mu_N\beta_i^2}{2\hbar^2}U_{NA,i}(r)G_i(r,E)\right)
\eeqn{u1p}
and 
\beq
G_i(r,E) = \exp\left[-\frac{\mu_N\beta_i^2}{2\hbar^2}(E - U_{loc}(r))\right].
\eeqn{Gi}

\begin{center}
\begin{figure*}[t]
\includegraphics[width=0.9\textwidth]{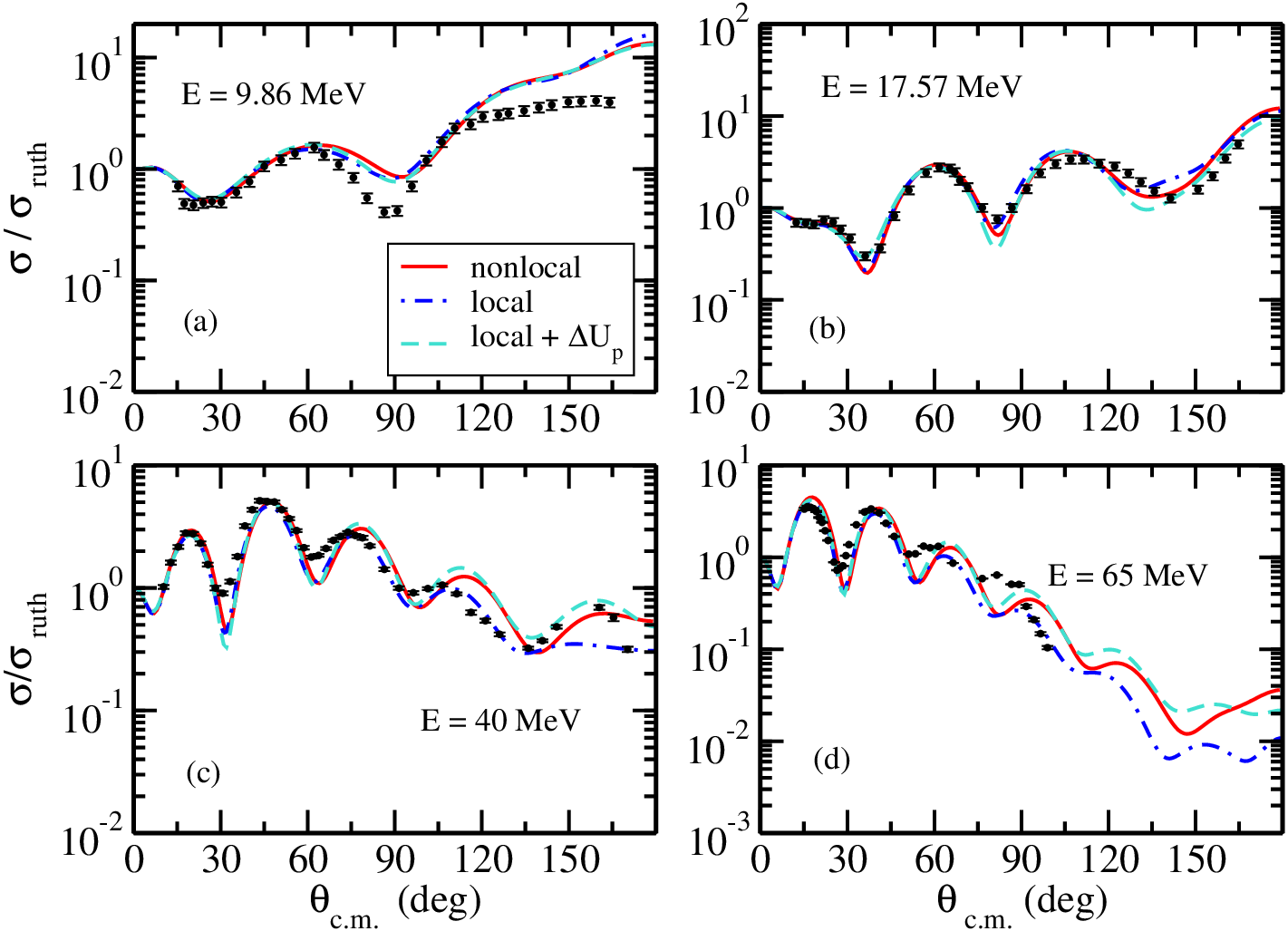}
\caption{(color online). Differential cross sections normalized by the Rutherford cross section for proton scattering on $^{40}$Ca at (a) 9.86 MeV, (b) 17.57 MeV, (c) 40 MeV and (d) 65 MeV, calculated using the fully nonlocal DOM potential (solid), using the local potential $U_{loc}$ from Eq.~(\ref{uloc_multi}) with the Coulomb and spin-orbit potentials included (dot-dashed), and using this equivalent local potential but with the correction $\Delta U_p$ included as well (dashed). These are compared with the experimental data (dots). }
\label{Fig:pca40_equiv}
\end{figure*}
\end{center}

 Figure~\ref{Fig:pca40_equiv} compares proton scattering differential cross sections (normalized by the Rutherford cross section) determined from solving the NLDOM scattering problem exactly using the iterative procedure outlined in Ref. ~\cite{Michel09} and from solving the local-equivalent problem. Both the Coulomb and spin-orbit corrections are included. The experimental data are from Refs. \cite{Dicello, McCamis, Blumberg, Noro}. Aside from small deviations at large angles, the results from using Eq.~(\ref{uloc_multi}) are very similar to the exact solutions. The results from including $\Delta U_p$ are also shown. Overall, this correction improves the correspondence between the exact and approximate solutions of the nonlocal problem for angles $ \theta \approx 40^\circ $ and above.

\begin{figure}
\includegraphics*[width=0.4\textwidth]{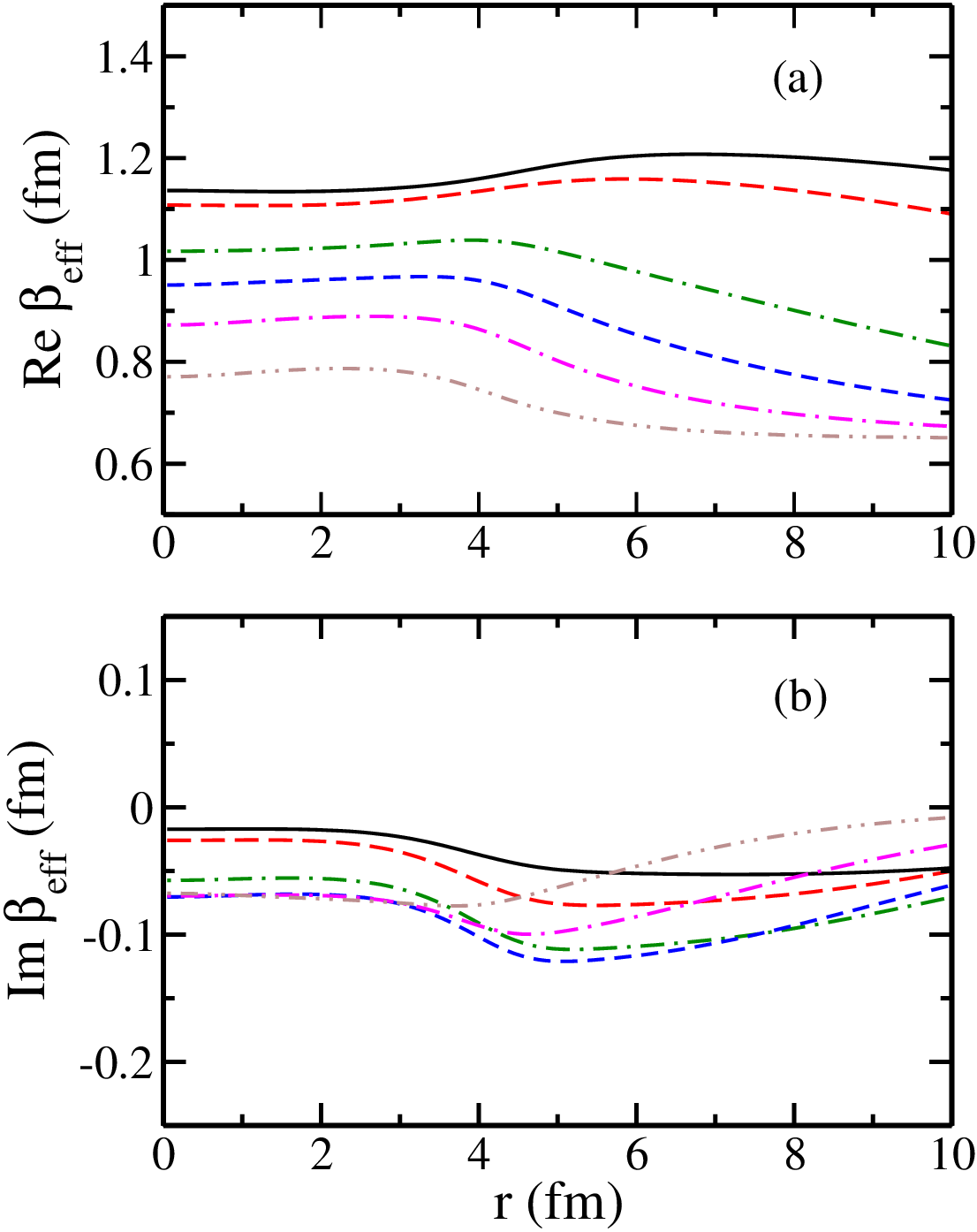}
\caption{(color online). (a) Real and (b) imaginary parts of $\beta_{\text{eff}}$ for E = 9.86 MeV (solid), 17.57 MeV (long dashed), 40 MeV (long dot-dashed), 65 MeV (short dashed), 100 MeV (short dot-dashed), and 200 MeV (dot-dot-dashed).}
\label{Fig:pbeff}
\end{figure}

For Perey-Buck potentials with one nonlocality parameter the $N-A$  wave function obtained from the phase-equivalent local model defined by a potential $U(r)=U_{loc}+\Delta U_N$ differs in the nuclear interior from the exact $N-A$ wave function by the Perey factor \cite{Fid}
\beq
f(r)=\exp \left[\frac{\mu_N \beta^2}{4 \hbar^2} U_{loc}(r)\right].
\eeqn{Perey_1}
Elastic scattering observables do not depend on the Perey factor. Transfer cross sections may depend on it if they are not peripheral. In the particular case of $^{40}$Ca(d,p)$^{41}$Ca, the internal part contributes up to 20$\%$ \cite{Pan07} for the energy being considered, and this contribution is more important in the DWBA  than in the ADWA. One can show it is also possible to derive the Perey factor for optical potentials with multiple nonlocalities such as in NLDOM. In this case the Perey factor is
\beq
f(r) =\exp \left[ \frac{\mu_N \beta_{\rm eff}^2(r)}{4 \hbar^2} U_{loc} \right],
\eeqn{Perey_2}
where
\beq
\beta_{\rm eff}^2(r)= -U_{loc}^{-1}(r)
\int_r^{\infty} dr_1 \frac{\sum_{i}\beta_i^2 U^\prime_{NA, i}(r_1)G_i(r_1, E)}{1 - U_1( r_1, E)}.
\eol
\eeqn{beff}
This Perey factor has some effective $r$-dependent range $\beta_{\rm eff}$, which can be complex. The real and imaginary parts are shown in Fig.~\ref{Fig:pbeff} for the case of p - $^{40}$Ca elastic scattering at several proton energies. The imaginary part is small and has a negligible effect on the (d,p) cross sections. We note that $\beta_{\text{eff}}$ decreases with energy. This decrease reflects the fact that $\beta_{sur+}$ is larger than $\beta_{vol+}$. Since the volume imaginary potential dominates at higher energies, the term in Eq.~(\ref{uloc_multi}) with $\beta_i = \beta_{vol+}$ becomes more important with increasing energy. This term also seems to dominate at large $r$, as Re $\beta_{\rm eff}$ converges to $\beta_{vol+} = 0.64$ fm for all energies. 

\begin{figure}
\includegraphics*[width=0.4\textwidth]{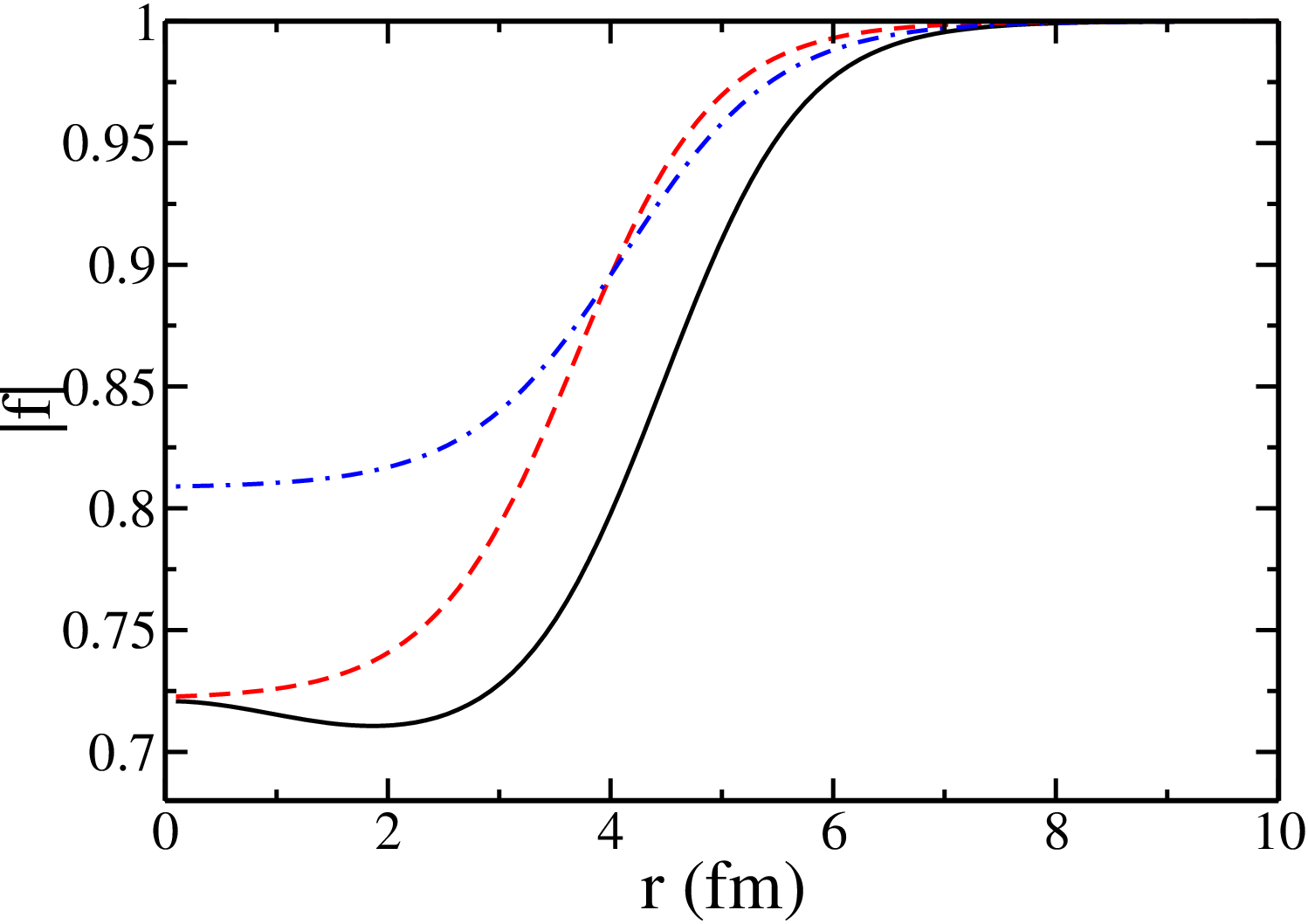}
\caption{(color online). Absolute value of the Perey factor evaluated at $E_{cm} = 17.37$ MeV for NLDOM (solid), LDOM (dashed) and CH89 (dot-dashed).}
\label{Fig:perey_l}
\end{figure}

The Perey factor for NLDOM is shown in Fig.~\ref{Fig:perey_l}, evaluated at the energy $E_p = 17.37$ MeV, which is the center of mass energy for the outgoing proton in the $^{40}$Ca(d,p)$^{41}$Ca reaction with $E_d = 11.8$ MeV (in the lab frame). It is compared to the Perey factor of an earlier version of the DOM \cite{Mueller11} that is purely local (LDOM). This Perey factor will be used in Sec.~\ref{sec:results}. It is given by \cite{Mahaux87, Mahaux91}

\beq
f(r,E) = \sqrt{\tilde{m}(r,E)/m},
\eeqn{fldom}
where $\tilde{m}(r,E)/m$ is the so-called momentum-dependent effective mass and is related to the LDOM Hartree-Fock potential as

\beq
\frac{\tilde{m}(r,E)}{m} = 1 - \frac{d{\cal V}_{HF}(r,E)}{dE}.
\eeqn{meff}
Figure~\ref{Fig:perey_l} also shows the Perey factor calculated with the widely used CH89 potential using Eq.(~\ref{Perey_1}) and assuming $\beta = 0.85$ fm. The Perey factors from LDOM and CH89 both have less effect in the surface region than the one calculated with NLDOM. 

To calculate the $^{40}$Ca(d,p)$^{41}$Ca cross sections a choice needs to be made for the optical potential $U_p$ in the exit channel. In principle, this potential is auxiliary, and it is believed that choosing $U_p$ that describes proton elastic scattering in the exit channel makes 
the remnant term $\sum_i V_{pi} - U_p$ in the transfer amplitude to disappear \cite{Satchler}. Since NLDOM was not fit to $p+^{41}$Ca scattering data, one choice for the auxiliary $p+^{41}$Ca potential $U_p$ is to use NLDOM but evaluated with $A = 41$ instead of $A = 40$. An alternative, originally proposed in \cite{GW} and then further explored in \cite{Tim99}, stems from the argument that the remnant term can be removed from the transition operator exactly, leading to a different model for the exit state wave function. In this model, the three-body Hamiltonian associated with the exit channel contains the  $p-^{40}$Ca optical potential, the $n-^{40}$Ca bound-state potential and no $n-p$ interaction. 
In the limit of infinitely large core and in the zero-range approximation, the corresponding three-body wave function contains the $p-^{41}$Ca distorted wave function calculated with the $p-^{40}$Ca optical potential. Corrections due to recoil excitation and breakup are considered in \cite{Tim99}. For light nuclei the validity of the transfer amplitude with no remnant has also been confirmed by \cite{Mor09}. We analyzed the (d,p) reaction with both choices for $U_p$, and the resulting cross sections were found to differ at the peak by about 1\%. For the purposes of this study, both choices give practically the same result. Below, we choose to use NLDOM evaluated with $A = 41$ for $U_p$.

\section{The deuteron-target potential for $(d,p)$ reactions in the adiabatic approximation}
\label{sec:adwa}

Following Johnson-Tandy \cite{JT} we retain only the first Weinberg component of the $A + n + p$ system in the $(d,p)$  transfer amplitude since this amplitude is sensitive  only to those parts of this wave function in which the neutron $n$ is close to the proton $p$. Recently,  exact continuum-discretized coupled channel calculations confirmed that this component indeed dominates  \cite{Pan13}. The first Weinberg component is a product of the deuteron wave function $\phi_0(\ve{r})$ times the $d-A$ relative motion wave function $\chi(\ve{R})$ which is the solution of the two-body Schr\"odinger equation with an adiabatic potential constructed from $p-A$ and $n-A$ optical potentials. The generalization of the deuteron adiabatic potential for the case of nonlocal, energy-independent $N-A$ optical potentials of the Perey-Buck type is given in \cite{Tim13b}. If nonlocal potentials (such as NLDOM) explicitly depend on energy then they should be evaluated at the energy $E=E_d/2+ \la T_{np}\ra_{V_{np}}/2$ and then treated as nonlocal and energy-independent \cite{Joh14}. The $\la T_{np}\ra_{V_{np}}/2$ term is half the $n-p$ kinetic energy in deuteron averaged over the short-ranged potential $V_{np}$. The value of this term is about 57 MeV \cite{Joh14}, so for $E_d = 11.8$ MeV, the NLDOM potential should be evaluated at about $63$ MeV. This at first sight seems counterintuitive. However, keeping in mind that neutron transfer takes place when proton and neutron in deuteron are at very short separations where the Heisenberg principle dictates high relative $n-p$ momentum, it becomes clear that there is an additional kinetic energy in the $N-A$ system which should be taken into account when choosing the energy at which the $N-A$ potential should be evaluated.

\subsection{Lowest order equivalent local model}
\label{sec:adwaA}

The nonlocal Schr\"odinger equation for  $\chi(\ve{R})$ from \cite{Tim13b} can easily be generalized
for the case of nonlocal optical potentials with multiple nonlocalities:
\beq
(T_R+ U_C(R)-E_d)\chi(\ve{R}) = -\sum_{N=n,p} \sum_i
 \int d\ve{s}\,d\ve{x} \eol \times \phi_1(\ve{x}+ \alpha_1\ve{s} ) 
 U_{NA,i}(\frac{\ve{x}}{2} - \ve{R})H_i(s)   
\phi_0(\ve{x}) \chi( \frac{\alpha_2\ve{s}}{2}+\ve{R})  \eol 
\eeqn{NLeq}
where  $\ve{R}$ is the radius-vector between $d$ and $A$, $T_R$ is the kinetic energy operator associated with $\ve{R}$, $\alpha_1 = A/(A+1)$, $\alpha_2 = (A+2)/(A+1)$, $\phi_0$ is the deuteron ground state wave function and
\beq
\phi_1(\ve{r})=\frac{V_{np}(\ve{r})\phi_0(\ve{r})}{\la\phi_0|V_{np}|\phi_0\ra}.
\eeqn{phi1}

\begin{table*}
\caption{ Coefficients $\beta_n^{(\lambda)}$ (in fm) and 
moments $M_{0 }^{(\lambda)}$  (in fm$^{2\lambda}$) for $\lambda = 0,1$ and for six different values of the nucleon nonlocality $\beta$ that are used in the NLDOM.}
\label{Tbl:betas}
\begin{tabular}{P{1.0 cm}P{2.0 cm}P{2.0 cm}P{2.0 cm}P{2.0 cm}P{2.0 cm}P{2.0 cm}}
\hline
\hline

 $n$ & $\beta = 0.64$ fm & $\beta = 0.84$ fm & $\beta = 0.94$ fm & $\beta = 1.04$ fm & $\beta = 1.55$ fm & $\beta = 2.07$ fm \\ \hline \\
$\beta^{(0)}_{1}$ & 0.3075 & 0.3932 & 0.4393 & 0.4802 & 0.6821 & 0.8679  \\
$\beta^{(0)}_{2}$ & 0.3082 & 0.3947 & 0.4414 & 0.4829 & 0.6895 & 0.8824  \\
$\beta^{(0)}_{3}$ & 0.3088 & 0.3960 & 0.4431 & 0.4851 & 0.6955 & 0.8938  \\
$\beta^{(0)}_{4}$ & 0.3093 & 0.3970 & 0.4445 & 0.4869 & 0.7004 & 0.9031  \\
$\beta^{(0)}_{5}$ & 0.3097 & 0.3979 & 0.4457 & 0.4885 & 0.7045 & 0.9108  \\
$\beta^{(0)}_{6}$ & 0.3101 & 0.3986 & 0.4468 & 0.4898 & 0.7080 & 0.9174  \\
$\beta^{(0)}_{7}$ & 0.3104 & 0.3993 & 0.4477 & 0.4910 & 0.7111 & 0.9231 \\
$\beta^{(0)}_{8}$ & 0.3107 & 0.3999 & 0.4486 & 0.4921 & 0.7139 & 0.9281  \\ \\

$M_{0}^{(0)}$ & 0.855 & 0.791 & 0.756 & 0.726 & 0.580 & 0.464  \\ \\

$\beta^{(1)}_{1}$ & 0.3100 & 0.3981 & 0.4459 & 0.4885 & 0.7030 & 0.9062  \\
$\beta^{(1)}_{2}$ & 0.3104 & 0.3991 & 0.4473 & 0.4903 & 0.7077 & 0.9151  \\
$\beta^{(1)}_{3}$ & 0.3108 & 0.3999 & 0.4484 & 0.4917 & 0.7115 & 0.9224  \\
$\beta^{(1)}_{4}$ & 0.3111 & 0.4006 & 0.4494 & 0.4930 & 0.7148 & 0.9286  \\
$\beta^{(1)}_{5}$ & 0.3115 & 0.4013 & 0.4502 & 0.4941 & 0.7177 & 0.9339  \\
$\beta^{(1)}_{6}$ & 0.3118 & 0.4018 & 0.4510 & 0.4951 & 0.7202 & 0.9384  \\
$\beta^{(1)}_{7}$ & 0.3120 & 0.4023 & 0.4517 & 0.4959 & 0.7224 & 0.9424  \\
$\beta^{(1)}_{8}$ & 0.3122 & 0.4028 & 0.4523 & 0.4967 & 0.7243 & 0.9459 \\ \\

$M_{0}^{(1)}$ & 0.098 & 0.149 & 0.179 & 0.206 & 0.341 & 0.451  \\ \\
\hline
\hline
\end{tabular}
\end{table*} 
Solving the nonlocal problem (\ref{NLeq}) directly is certainly possible, and has recently been done in Ref.~\cite{Titus16}. However, in this paper, we construct the local-equivalent model, as simplified local-equivalent models can provide useful insight into the physics of a problem and make available transfer reaction codes applicable to nonlocal problems. The local-equivalent approximation of (\ref{NLeq})  can be obtained by expanding both $U_{NA,i}(\frac{\ve{x}}{2}-\ve{R})$ and $\chi( \frac{\alpha_2\ve{s}}{2}+\ve{R}) $ into Taylor series. In the lowest order approximation, using $U_{NA,i}(\frac{\ve{x}}{2}-\ve{R})\approx U_{NA,i}(\ve{R}) $ we get
\beq
(T_{R}+U_C(R)-E_d)\chi(\ve{R}) =
-  \sum_i{U_{dA, i}(R){\tilde H}^{(0)}_i(T_R)}\chi(\ve{R}), \eol \,\,\,\,\,\,\,\,\,\,\,
\eeqn{LM0}
where $U_{dA, i}(R)= U_{nA, i}(R)+U_{pA, i}(R)$,
\beq
\tilde{H}^{(0)}_i(T_R) = M^{(0)}_{0,i}\sum_{n=0}^{\infty} \frac{(-)^n}{n!} \left(\frac{\mu_d\alpha_2^2}
{2\hbar^2}(\beta_{n,i}^{(0)})^2T_R\right)^n, 
 \,\,\,\,\,\,
\eeqn{expandH0}
the coefficients $\beta_{n,i}^{(0)}$ are defined by
\beq
 \beta_{n, i}^{(0)} =\frac{1}{\sqrt{2}}\left[\frac{M^{(0)}_{2n, i}}{(2n+1)!! M^{(0)}_{0, i}}\right]^{\frac{1}{2n}}, \,\,\,\,\,\,\,\,
\eeqn{betan}
and the moments $M^{(0)}_{2n, i}$ are defined by  
\beq
M^{(0)}_{2n, i} = \int d\ve{s}d\ve{x} \, s^{2n } H_i(s) \phi_1(\ve{x}-\alpha_1\ve{s}) \phi_0(\ve{x}). 
\eeqn{M02n}

Eq. (\ref{LM0}) is further simplified by introducing the local-energy approximation 
\cite{Satchler},
\beq
T_R \approx T_0(R) = E_d - U_{loc}^0(R) - U_C(R) ,
\eeqn{K2}
where the local potential $U_{loc}^0(R)$ is defined as
\beq
U_{loc}^0(R) =  \sum_i{U_{dA, i}(R){\tilde H}_i^{(0)}(T_0(R))}.
\eeqn{ul}
This approximation works very well for nucleon optical potentials with one nonlocality parameter \cite{Tim13b}. We will show that it remains good for NLDOM  with its multiple nonlocalities, but we first present the results of calculations of $U_{loc}^0(R)$ for $E_d$ = 11.8 MeV using the deuteron wave function from the Hult\`en model, the same as in \cite{Tim13b}. It is pointed out in \cite{Tim13b} that a realistic deuteron wave function gives the moment $M^{(0)}_0$ which is very similar to that obtained with the Hult\'en wave function. In Table~\ref{Tbl:betas} we show the calculated $\beta_{n,i}^{(0)}$ and $ M^{(0)}_{0,i}$ terms. For very small nonlocalities, $\beta_{n,i}^{(0)}$ should be approximately equal to $\beta_{i}/2$ \cite{Tim13a,Tim13b}. For typical nonlocalities of $\sim 0.8-1.0$ fm they are $\sim 10\%$ smaller than the $\beta_{i}/2$ value but are essentially independent of $n$, allowing one to replace the $\beta_{n,i}^{(0)}$ coefficients with a constant. We define this constant to be 

\beq
\beta_{d,i} = \beta_{1,i}^{(0)}.
\eeqn{betadi}

The summation in (\ref{expandH0}) then gives $\tilde{H}_i^{(0)}(T_R)$ an exponential form
\beq
\tilde{H}_i^{(0)}(T_R)\approx {\cal H}_i^{(0)}(\gamma_i T_R) = M_{0,i}^{(0)} \exp\left(-\gamma_i T_R\right),
\eeqn{H0}
where
\beq
\gamma_i = \frac{\mu_d\alpha_2^2\beta_{d,i}^2}{2\hbar^2}.
\eeqn{gamma}

\begin{figure}
 \includegraphics*[width=0.4\textwidth]{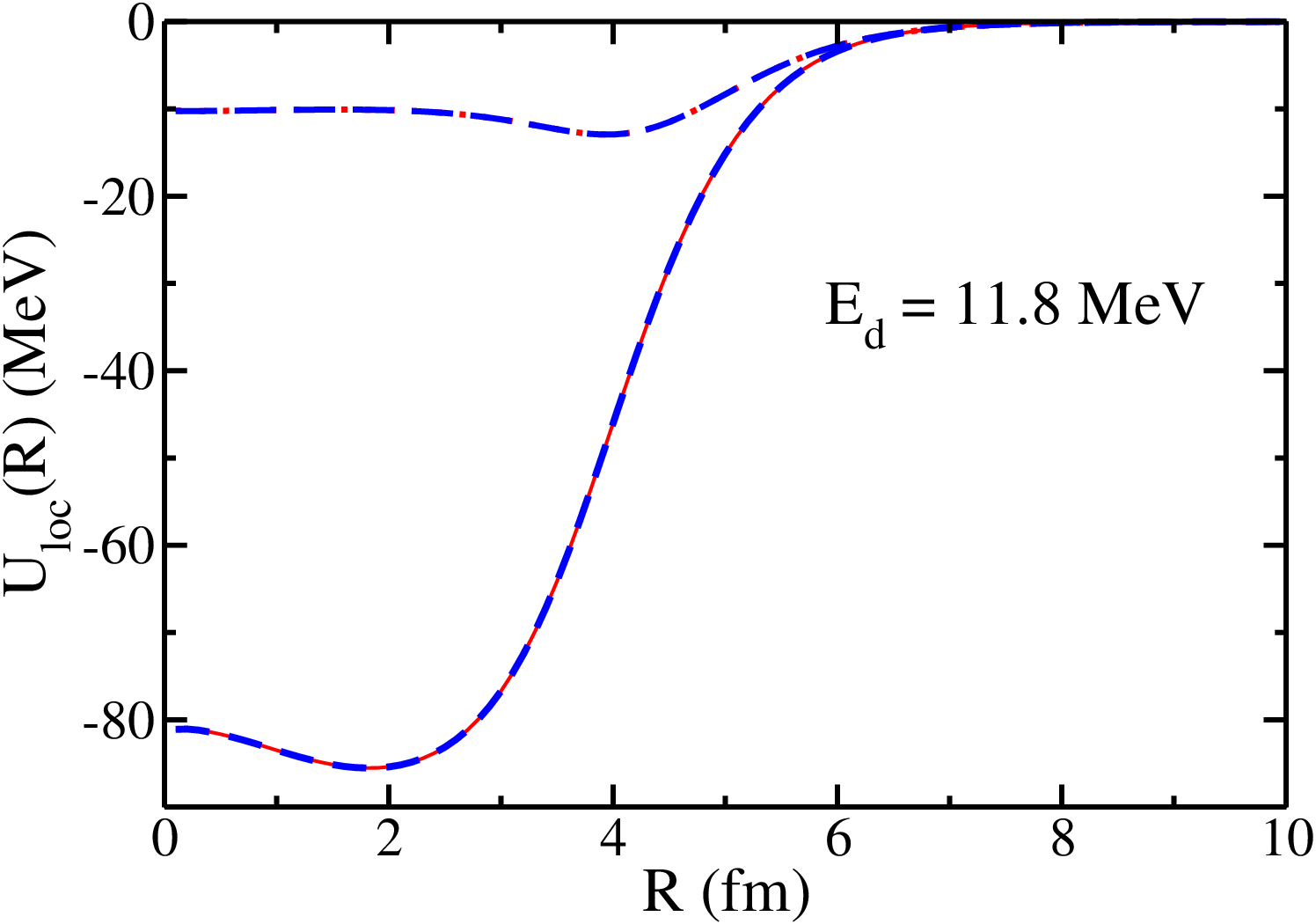}
 \caption{ (color online). Real and imaginary parts of the local-equivalent NLDOM potential calculated using the exponential form (solid and dotted lines, respectively) and using the series form (thick short-dashed and long-dashed lines, respectively) with $n_{max} = 6$.}
 \label{Fig:series}
 \end{figure}

As $\beta_i$ increases beyond 1.0 fm, the approximation in Eq.~(\ref{betadi}) becomes less valid and the coefficients $\beta_{n,i}^{(0)}$ deviate from $\beta_i/2$ even more. This trend can be seen in Table~\ref{Tbl:betas} and is especially apparent for $\beta = 2.07$ fm, which is the largest nonlocality parameter used in NLDOM. Thus, we solve Eqs. (\ref{expandH0}), (\ref{K2}), (\ref{ul}) by restricting the sum over $n$ in (\ref{expandH0}) by some $n_{\max}$. We found that $n_{\max}=6$ is sufficient to obtain a converged solution for $U_{loc}$. However, we also found that using Eq.~(\ref{betadi}) leads to practically the same result because the real and imaginary parts of the NLDOM terms associated with $\beta = 2.07$ fm are small at the energies being considered. The $U_{loc}$ potentials obtained from solution of (\ref{ul}) with (\ref{expandH0}) and (\ref{H0}) are shown in Fig.~\ref{Fig:series}. Using Eq.~(\ref{betadi}) for all nonlocality parameters, one obtains $U_{loc}$ from the transcendental equation
\beq
U_{loc}^0 &= &\sum_i{M_{0, i}^{(0)}U_{dA, i}} \times \nonumber \\
 &&\exp\left[-\frac{ \mu_d\alpha_2^2\beta_{d,i}^2} {2\hbar^2} (E_d-U_{loc}^0-U_C)\right]. \,\,\,\,
\eeqn{ulocd}
The Coulomb potential $U_C$ is approximated by a constant, given by
\beq
U_C = -1.08 + \frac{1.35 Z}{A^{1/3}},
\eeqn{uc}
which was used in Refs.~\cite{Gia76, GRZ}. The difference between using this approximation and a more realistic potential is only about 1\%, in terms of the peak cross section of the proton angular distribution for the $^{40}$Ca(d,p)$^{41}$Ca reaction at $E_d = 11.8$ MeV. 

\subsection{Correction to the local-energy approximation in the lowest order}

It was shown in \cite{Tim13b} that corrections to the lowest order local model beyond the local-energy approximation are  small because they are the fourth-order effect of the nucleon nonlocality $\beta$, as the second-order terms cancel each other for Perey-Buck potentials with one nonlocality parameter. The NLDOM from \cite{NLDOM} contains several nonlocalities, and second-order contributions may not cancel. Moreover, some of these nonlocalities are large so that the contributions beyond the local-energy approximation are expected to be larger than those in \cite{Tim13b}. In this section we study these corrections using results from sections IV.C and A.4  of \cite{Tim13b}. Including leading correction term, linear in the kinetic energy operator $T_R$, and using the exponential form (\ref{H0}) for $\tilde{H}_i^{(0)}(T_R)$, the right-hand side of Eq.~(\ref{LM0}) becomes

\beq
\sum_i U_{dA,i}{\cal H}_i^{(0)}(\gamma_i T_R)\chi(\ve{R})  
\approx \sum_i U_{dA,i}{\cal H}_i^{(0)}(\gamma_i T_0) 
\eol \times
\biggl[1-\gamma_i(T_R - T_0 + \Delta_i) 
- \frac{\hbar^2\gamma_i^2}{2\mu_d}\ve{\nabla}T_0\cdot\ve{\nabla}_R\biggr]\chi(\ve{R})\,\,\,\,\,\,\,\,,
\eeqn{loc_corr} 
where the energy correction $\Delta_i$, arising because $T_R$ and $T_0$ do not commute, is given by
\beq
\Delta_i = \frac{\hbar^2\gamma_i}{2\mu_d}\left(\frac{T_0''}{2} + \frac{T_0'}{R} - \frac{\gamma_i}{3}T_0'^2\right).
\eeqn{Delta}
The solution of Eq. (\ref{LM0}) in this approximation is the product 
\beq
\chi(\ve{R}) = f_0(\ve{R})\varphi(\ve{R}), 
\eeqn{}
where $\varphi$ is the scattering wave of the local model
\beq
( T_R + U_C-E_d)\varphi=   - (U^0_{loc} + \Delta U_0)\,\varphi.
\eeqn{cor1}
The $U^0_{loc}$ term is discussed in the previous section, and   the correction term $\Delta U_0$ is
\beq
\Delta U_0 = \frac{T_R f_0}{f_0} + \frac{\hbar^2}{4\mu_d}\left(\frac{U_2 T_0'}{1 - U_1}\right)^2 - \frac{U_{\Delta}}{1- U_1}
\eeqn{delU0} with
\beq
U_n(R) = \sum_i U_{dA,i}{\cal H}_i^{(0)}(\gamma_i T_0)\gamma_i^n, \\
U_{\Delta}(R) = \sum_i U_{dA,i}{\cal H}_i^{(0)}(\gamma_i T_0)\gamma_i \Delta_i.
\eeqn{U123}

The function  $f_0$ is the analog of the Perey factor discussed in Sec. \ref{sec:nldom}. It modifies the scattering wave function $\varphi(\ve{R})$ in the nuclear interior and satisfies the first order differential equation
\beq
    \frac{\ve{\nabla}f_0}{f_0}= -\frac{1}{2} \,\frac{ U_{2}(R)}{1-U_{1}(R)} \ve{\nabla}  T_0 ,
\eeqn{dff0} 
with the boundary condition $f(R) \rightarrow 1$ at $R\rightarrow \infty$. 
The solution of this equation is
\beq
 f_0(R) = \exp \left(\frac{1}{2} \int_R^{\infty} dR_1\frac{ U_{2}(R_1)}{1-U_{1}(R_1)}T'_0(R_1)\right).
\eeqn{f0}
Because of multiple nonlocalities, the analytical integration in (\ref{f0}) cannot be done. So, it is difficult to see if the contributions to $f_0$ from  second-order terms  on $\beta_{d,i}$ cancel. Most likely, they do not. 

\begin{figure}
\includegraphics*[width=0.4\textwidth]{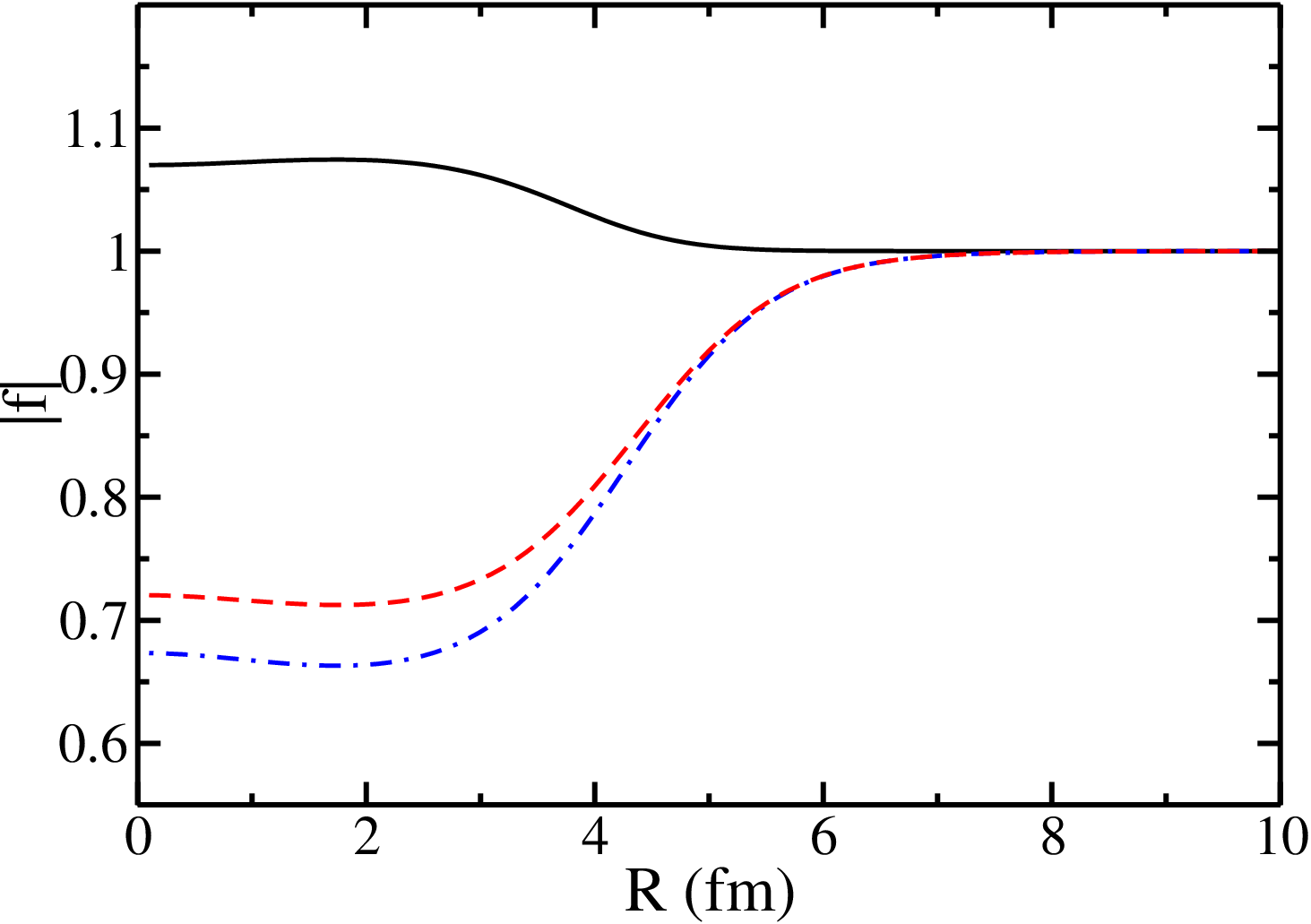}
\caption{(color online). Perey factors $f_0$ (solid), $f_1$ (dot-dashed), and $f$ (dashed) calculated with NLDOM for $E_d = 11.8$ MeV.}
\label{Fig:f_dom}
\end{figure}

\begin{figure}
\includegraphics*[width=0.4\textwidth]{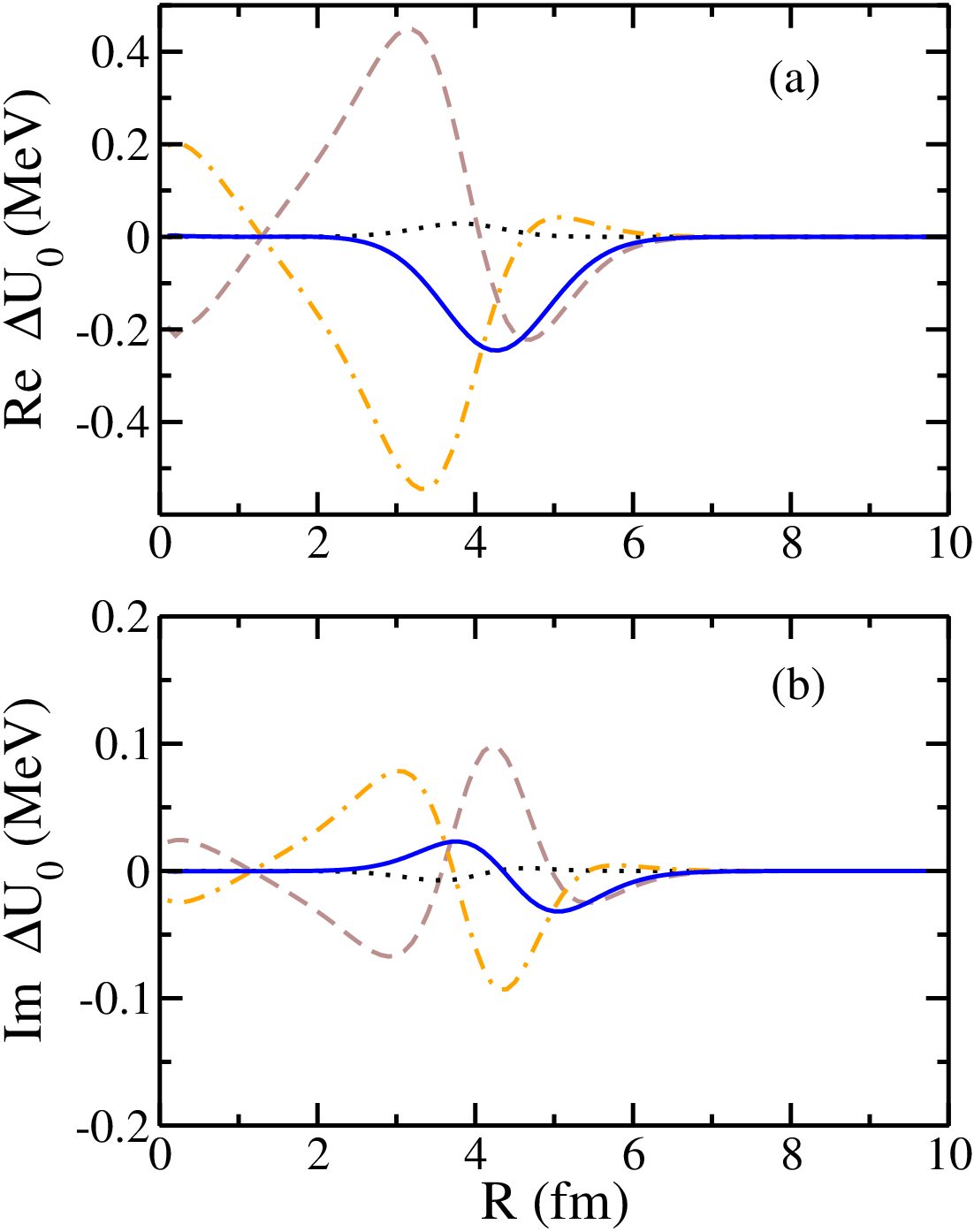}
\caption{(color online). (a) The real parts and (b) the imaginary parts of $\Delta U_0$ (solid) and the first (dashed), second (dotted), and third (dot-dashed) terms in Eq.~(\ref{delU0}).}
\label{Fig:delU0}
\end{figure}

The Perey factor $f_0$ and the correction $\Delta U_0$ to the equivalent local
potential $U_{loc}^0$ are shown in Fig.~\ref{Fig:f_dom} and Fig.~\ref{Fig:delU0}, respectively, for $d+^{40}$Ca at a deuteron 
incident energy of  11.8 MeV. The Perey factor
 increases the scattering wave in the nuclear interior by about 6$\%$, which is a couple of percent higher than the result in \cite{Tim13b}. The correction to $U_{loc}^0$, however, remains small.
 Its real part is very close to the one obtained in \cite{Tim13b} in the maximum, being about 150 keV, while the imaginary part is much smaller. Thus, for NLDOM the second order corrections most likely remain small and the local-energy approximation remains good.

\subsection{First order corrections} 

The first order correction to the local-equivalent lowest-order model of Sec.~\ref{sec:adwaA} is obtained by retaining two terms in the Taylor series expansion of the central potential $U_{NA}(\pm \frac{\ve{x}}{2}-\ve{R})$:
\beq
U( \pm \frac{\ve{x}}{2}-\ve{R}) \approx
U(\ve{R})\mp \frac{1}{2}\ve{x}\cdot \ve{\nabla} U(\ve{R}).
\eeqn{}
In this case, using techniques of \cite{Tim13b}, we obtain the following:

\beq
(T_{R}+U_C(R)-E_d)\chi(\ve{R}) =\qquad \qquad\qquad \qquad\qquad\qquad \eol
-\sum_i U_{dA,i}(R){\cal H}^{(0)}_i(\gamma_i T_R)\chi(\ve{R})\qquad\qquad\quad\eol
\quad-\sum_i\ve{\nabla}[U_{dA,i}(R)]{\cal H}^{(1)}_i(\tilde{\gamma}_i T_R)\ve{\nabla}\chi(\ve{R}), \,\,\,\,\,\,\,\,\,\,\,\,
\eeqn{LM2}
where 
\beq
{\cal H}^{(1)}_i(\tilde{\gamma}_i T_R)  = \frac{M^{(1)}_{0,i}}{3M^{(0)}_{0,i}} {\cal H}^{(0)}_i(\tilde{\gamma}_i T_R) 
\eeqn{H1}
and the moments $M^{(1)}_{2n,i}$ are defined as 
 \beq
M^{(1)}_{2n,i} = \int d\ve{s}d\ve{x} \,  s^{2n}H_i(s) 
\phi_1(\ve{x}-\alpha_1 \ve{s}) \phi_0(\ve{x})  \frac{\alpha_2\ve{s}\cdot\ve{x}}{4}
  . \,\,\,\,\,\,\,\,
\eeqn{M12n}
The new factor $\tilde{\gamma}_i$ arises from the fact that the moments $M^{(1)}_{2n,i}$ lead to new coefficients $\beta^{(1)}_{n,i}$ that are also practically independent of $n$ (see Table~\ref{Tbl:betas}). Introducing a new constant
\beq
\tilde{\beta}_{d,i} = \beta^{(1)}_{1,i}
\eeqn{betadi2}
the factor $\tilde{\gamma}_i$ can be written as
\beq
\tilde{\gamma}_i = \frac{\mu_d\alpha_2^2\tilde{\beta}_{d,i}^2}{2\hbar^2}.
\eeqn{gamma2}
The coefficients $\beta^{(1)}_{n,i}$ are defined as
\beq
\beta^{(1)}_{n,i} = \frac{1}{\sqrt{2}}\left[\frac{3 M^{(1)}_{2n,i}}{(2n+3)!! M^{(1)}_{0,i}}\right]^{\frac{1}{2n}}.
\eeqn{beta1n} 

At this point we make the local-energy approximation (\ref{K2}) but we also include the correction to this approximation similar to that considered above. We expand  ${\cal H}^{(0)}_i(\gamma_i T_R)$ in the first term of r.h.s of Eq.~(\ref{LM2}) as in Eq.~(\ref{loc_corr}), but we use a simpler expansion for ${\cal H}^{(1)}_i(\tilde{\gamma}_i T_R)$ in the second term,
\beq
{\cal H}^{(1)}_i(\tilde{\gamma}_i T_R) ={\cal H}^{(1)}_i(\tilde{\gamma}_i T_0)( 1 - \tilde{\gamma}_i(T_R - T_0)),
\eeqn{H1_corr}
because ${\cal H}^{(1)}_i(\tilde{\gamma}_i T_R)$ is an order of magnitude smaller than ${\cal H}^{(0)}_i(\gamma_i T_R)$ so that the higher order terms on $\tilde{\beta}_{d,i}$ in terms with $\Delta_i$ and $\ve{\nabla}\cdot\ve{\nabla}_R$ will be small. 
For a similar reason we keep only the leading correction to $(T_R - T_0)\ve{\nabla}\chi(R)$:
\beq
(T_R - T_0)\ve{\nabla}\chi(R)\approx\ve{\nabla}T_0\, \chi(R).
\eeqn{TRmT0}
With these approximations we can solve Eq. (\ref{LM2}) by introducing the same representation 
$\chi(\ve{R}) = f(\ve{R})\varphi(\ve{R})$ used both for proton scattering in Sec. \ref{sec:nldom} and for correction to local-energy-approximation above. 
The  scattering wave $\varphi$ is found from the local equation
\beq
( T_R + U_C- E_d)\varphi=   -( U_{loc}^0 + \Delta U) \, \varphi ,
\eeqn{}
with the same $U_{loc}$ as before but corrected by
\beq
\Delta U = \Delta U_0 + \Delta U_1,
\eeqn{delU}
in which the first term is the same as in Eq.~(\ref{delU0}), and the second term is given by 
\beq
\Delta U_1 = \frac{T_R f_1}{f_1} + \frac{\mu_d}{\hbar^2}\frac{{\cal U}_0^2}{(1-U_1)^2} - \frac{{\cal U}_1 T_0'}{1-U_1} - \frac{1}{2}\frac{{\cal U}_0U_2T_0'}{(1-U_1)^2},\eol
\eeqn{delU1}
where
\beq
{\cal U}_n(R) &=& \sum_i\ve{\nabla}[U_{dA,i}(R)]{\cal H}^{(1)}_i(\tilde{\gamma}_i T_0)\tilde{\gamma}_i^n. 
\eeqn{u45}
The Perey factor $f$ is the solution of the first order differential equation
\beq
\frac{\ve{\nabla}f}{f}= \frac{\mu_d}{\hbar^2}\frac{{\cal U}_0}{1 - U_1} - \frac{1}{2}\frac{U_2 T_0'}{1-U_1}
\eeqn{dff}
with the boundary condition $f(R) \rightarrow 1$ at $R\rightarrow \infty$. The solution to this equation can be written as 
\beq
f(R) = f_0(R)f_1(R)
\eeqn{f}
where $f_0$ is given by Eq.~(\ref{f0}) and $f_1$ is given by
\beq
f_1(R) = \exp\left(- \frac{\mu_d}{\hbar^2}\int_R^\infty dR'\frac{{\cal U}_0(R') }{ 1 - U_1(R')}\right).
\eeqn{f1}

\begin{figure}
\includegraphics*[width=0.4\textwidth]{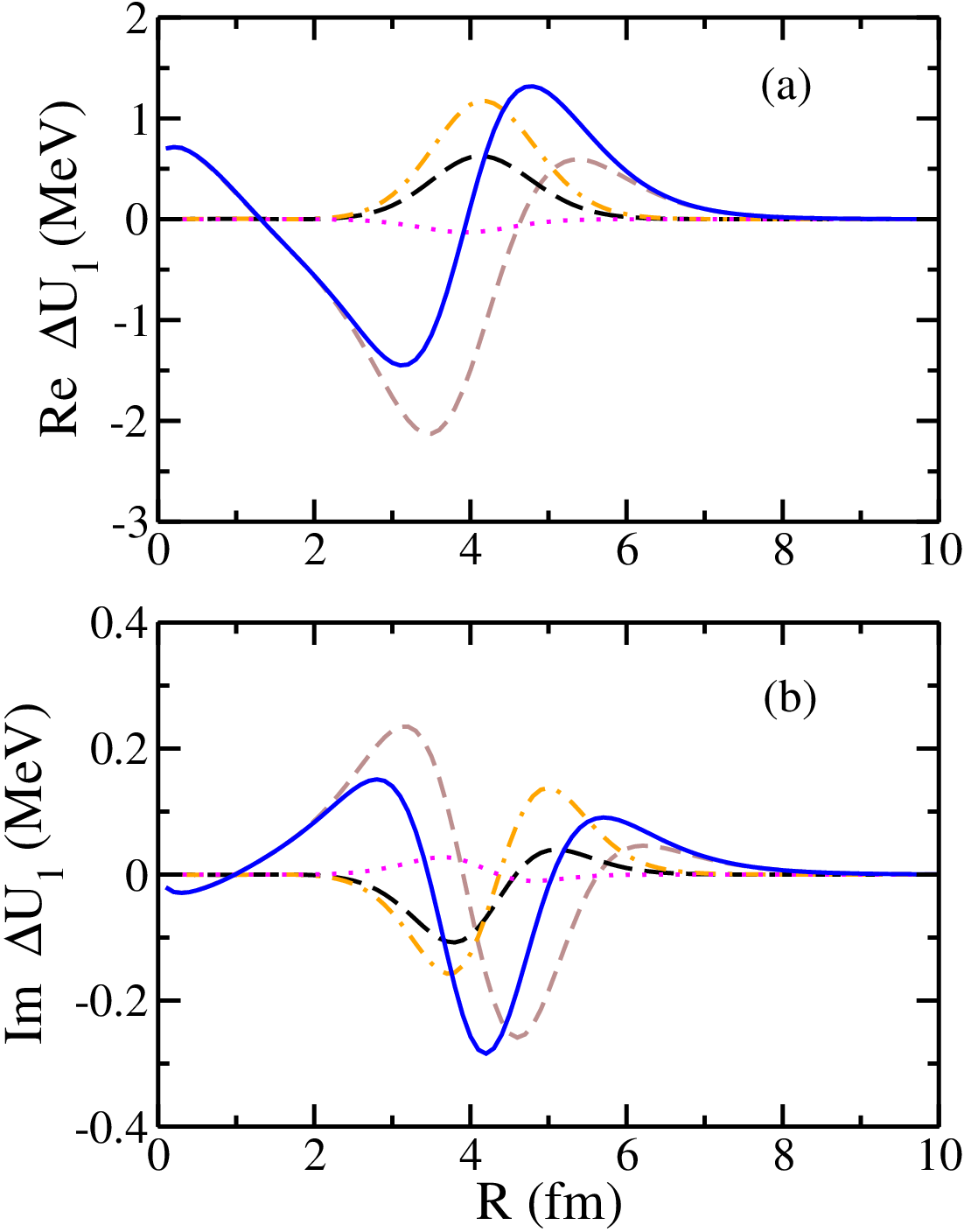}
\caption{(color online). (a) The real parts and (b) the imaginary parts of $\Delta U_1$ (solid) and the first (short-dashed), second (long-dashed), third (dot-dashed), and fourth (dotted) terms in Eq.~(\ref{delU1}).}
\label{Fig:delU1}
\end{figure}

\begin{figure}
\includegraphics*[width=0.4\textwidth]{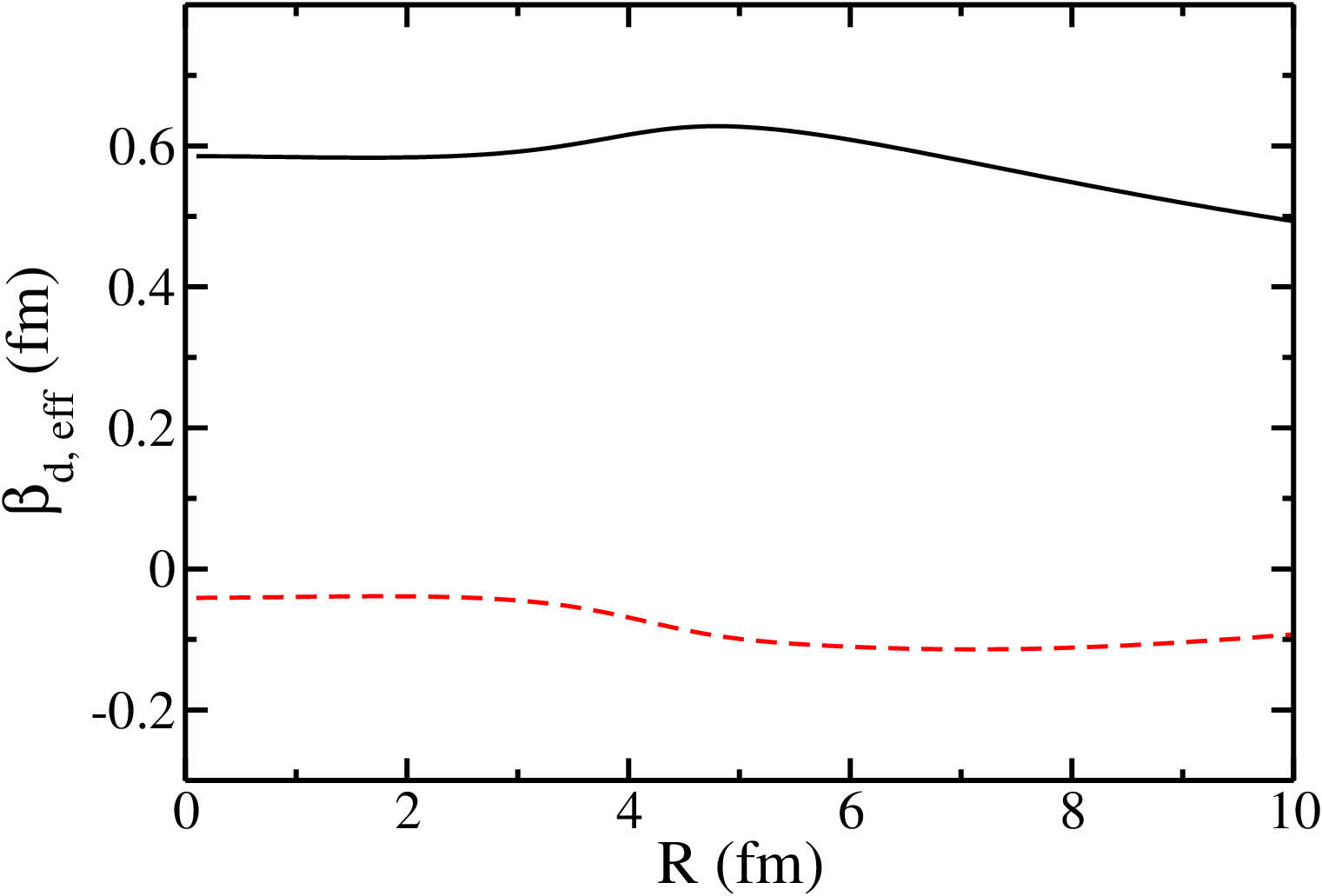}
\caption{(color online). Real part (solid) and imaginary part (dashed) of effective deuteron nonlocality as a function of $r$ for $E_d = 11.8$ MeV.}
\label{Fig:bdeff}
\end{figure}

The correction $\Delta U_1$ and the four terms on the right-hand side of Eq. (47) are shown in Fig.~\ref{Fig:delU1}. The Perey factors $f_1$ and $f$ are plotted in Fig.~\ref{Fig:f_dom}. The correction $\Delta U_1$ and the Perey factor $f$ are comparable to the corresponding quantities obtained in [7] for a Perey-Buck potential with a single nonlocality. We can rewrite the Perey factor $f$ in the form of Eq.~(\ref{Perey_2}). The corresponding effective deuteron nonlocality, $\beta_{d,\rm eff}$ is plotted in Fig.~\ref{Fig:bdeff} as a function of $r$ for $E_d = 11.8$ MeV. Note that $\beta_{d, \rm eff}$ is complex, but, as in the case for protons, the imaginary part is small and changes the cross section of the proton angular distribution by less than 0.5\%. The real part lies between 0.50 and 0.60 fm, and these values are very similar to 0.56 fm, which is the value used for deuteron elastic scattering.

\section{Transfer reaction $^{40}$C\MakeLowercase{a(d,p)}$^{41}$C\MakeLowercase{a} \MakeLowercase{at} 11.8 MeV}
\label{sec:results}

We calculated the proton angular distributions for the $^{40}$Ca(d,p)$^{41}$Ca reaction  at $E_d = 11.8$ MeV
in the adiabatic zero-range   approximation. The finite range effects at these energies are expected to be small \cite{FR}. The standard value for $D_0$, given by $D_0^2 = 15615$ MeV$^2$fm$^{3}$, was used. The distorted potentials both in the deuteron and proton channels, generated with NLDOM, were read into the TWOFNR code \cite{2FNR}.  There is no option in TWOFNR for incorporating complex $r$-dependent effective nonlocalities $\beta_{\rm eff}(r)$. Therefore, in order to reduce the corresponding distorted waves in the nuclear interior, we multiplied the NLDOM $\la ^{40}$Ca$|^{41}$Ca$\ra$ overlap function (also read into the TWOFNR code)  by the Perey factors of the proton and deuteron channels, given by Eqs. (\ref{Perey_2}) and (\ref{f}), respectively. This is legitimate in the zero-range approximation, where the integrand of the (d,p) reaction amplitude is a function of only one vector variable. In this case, the Perey factor for the proton channel had to be calculated on a different grid. 

\begin{figure}
\includegraphics*[width=0.4\textwidth]{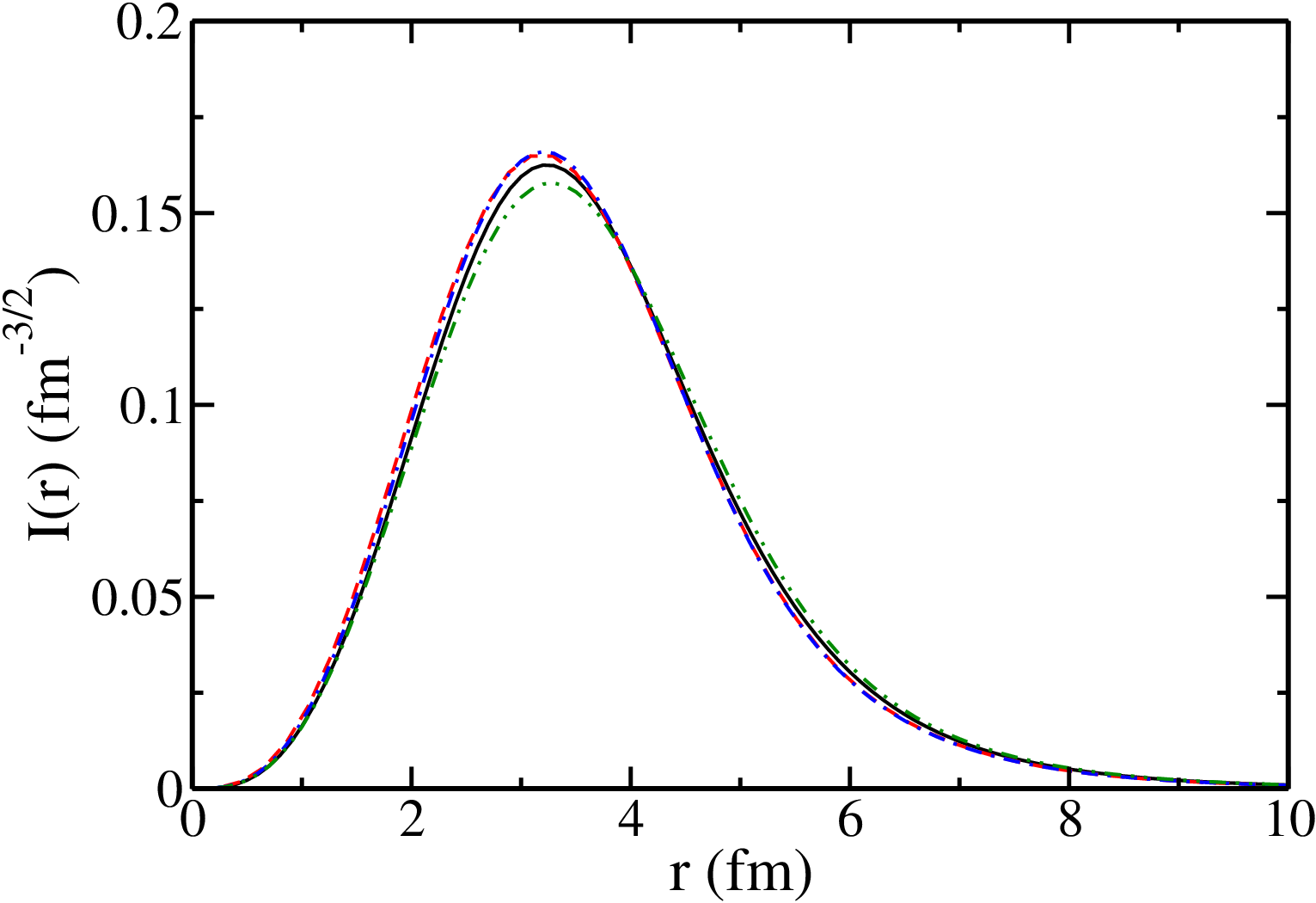}
\caption{(color online). Overlap functions calculated using NLDOM (solid), LDOM (dashed), a Woods-Saxon potential with standard geometry (dot-dashed), and a Woods-Saxon potential but corrected with a Perey factor with $\beta = 0.85$ fm (dot-dot-dashed).}
\label{Fig:ovl}
\end{figure}

The overlap function $I^{\rm NL}_{\rm DOM}(r)$, generated by NLDOM and read into TWOFNR, is compared in Fig.~\ref{Fig:ovl} to (i)  the overlap function $I_{\rm WS}(r)$ obtained from a Woods-Saxon potential with standard geometry ($r_0 = 1.25$ fm, $a = 0.65$ fm); (ii) the overlap function $I^L_{\rm DOM}(r)$ generated with LDOM and (iii) the overlap function $I^{NL}_{\rm WS}(r)$, calculated in a standard Woods-Saxon model employing a nonlocality correction via the Perey factor with $\beta = 0.85$ fm. All these overlap functions are normalized to 0.73, which is the spectroscopic factor calculated from NLDOM.

We have found that $I^{\rm NL}_{\rm DOM}(r)$ can be described very well (with about 1\% accuracy) by a local two-body Woods-Saxon potential model that has the radius $r_0=$ 1.252 fm, diffuseness $a = 0.718$ fm and the spin-orbit strength $V_{s.o.} = 6.25$ MeV. These parameters are very close to the standard values of $r_0$ = 1.25 fm and $a= $0.65 fm used to generate $I_{\rm WS}(r)$. However, relative to $I_{\rm WS}(r)$, using $I^{\rm NL}_{\rm DOM}(r)$ as the overlap function increases the transfer cross section at the peak, $\sigma_{d,p}^{peak}$, by about 15\% (for the reaction at $E_d = 11.8$ MeV).

The  NLDOM and LDOM overlap functions have similar shapes, but their r.m.s. radii and asymptotic normalization coefficients (ANC) somewhat differ. The $I^{\rm NL}_{\rm DOM}(r)$ radius of 4.030 fm is slightly larger than that of $I^{\rm L}_{\rm DOM}(r)$, which is 3.965 fm. Also, the   single-particle ANC $b_{\ell j}$ for $I^{\rm NL}_{\rm DOM}(r)$ is  $2.48$ fm$^{-1/2}$, which is about 10\% larger than that of $I^L_{\rm DOM}(r)$. The spectroscopic factors for these two overlaps are practically the same. As a result, $I^{\rm NL}_{\rm DOM}(r)$ produces a larger many-body ANC squared, $C_{\ell j}^2 = S_{\ell j}b_{\ell j}^2$, equal to $ 4.5$ fm$^{-1}$, whereas   $I^L_{\rm DOM}(r)$  has $C_{\ell j}^2 = 3.8$ fm$^{-1}$.
Interestingly, the NLDOM value of $C_{\ell j}^2$ is very close to the prediction of $C_{\ell j}^2 = 4.4$ fm$^{-1}$ of the source term approach \cite{Tim11},  which is based on the independent-particle-model for $^{40}$Ca and $^{41}$Ca. This approach accounts for correlations between nucleons  via an effective interaction potential of the removed nucleon with nucleons in the core \cite{Tim09}.

The standard overlap $I_{\rm WS}(r)$ is very close to $I^L_{\rm DOM}(r)$ (see Fig.~\ref{Fig:ovl}). The overlap $I^{NL}_{\rm WS}(r)$, which is sometimes  used in (d,p) calculations, has a distinctly larger radius and is not consistent with NLDOM. Below, in all our (d,p) calculations we use only the NLDOM overlap with its own normalization of 0.73, which allows for making conclusions from comparison between theoretical and experimental cross sections  without any further renormalization. 

\begin{figure}
\includegraphics*[width=0.4\textwidth]{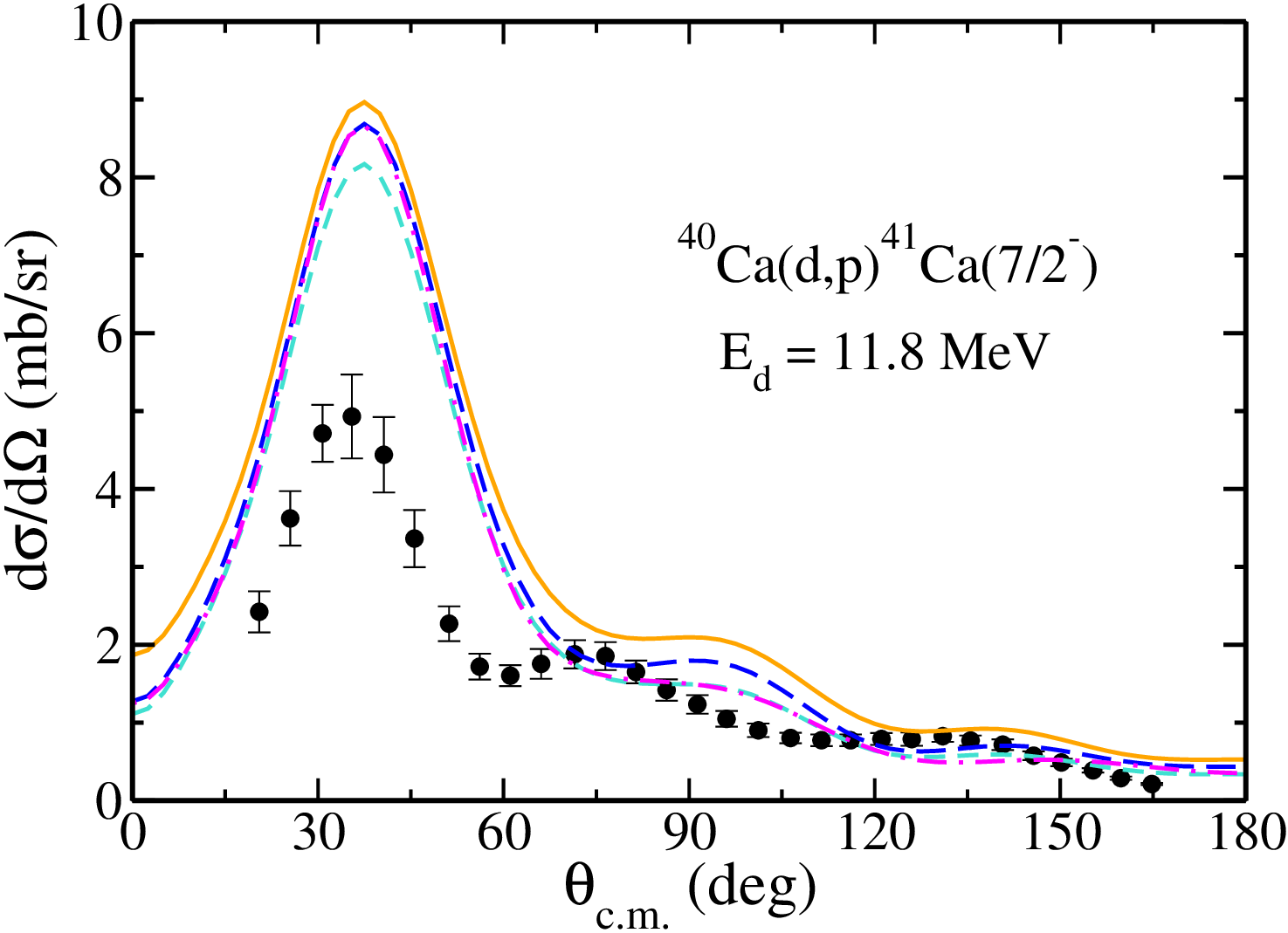}
\caption{(color online). Proton angular distributions for the $^{40}$Ca(d,p)$^{41}$Ca($7/2^-$) reaction with $E_d = 11.8$ MeV generated using NLDOM without the corrections from Eqs. (\ref{Delta_U}), (\ref{Perey_2}) and Eqs. ~(\ref{delU}), (\ref{f}) (solid), with the corrections for the proton channel (long dashed), with the corrections for both the proton and deuteron channels (short dashed), and with the spin-orbit potential in addition to the other corrections (dot-dashed). The experimental data are also shown (dots).}
\label{Fig:xsec_dom_mod}
\end{figure}

\begin{figure}[b]
\includegraphics*[width=0.4\textwidth]{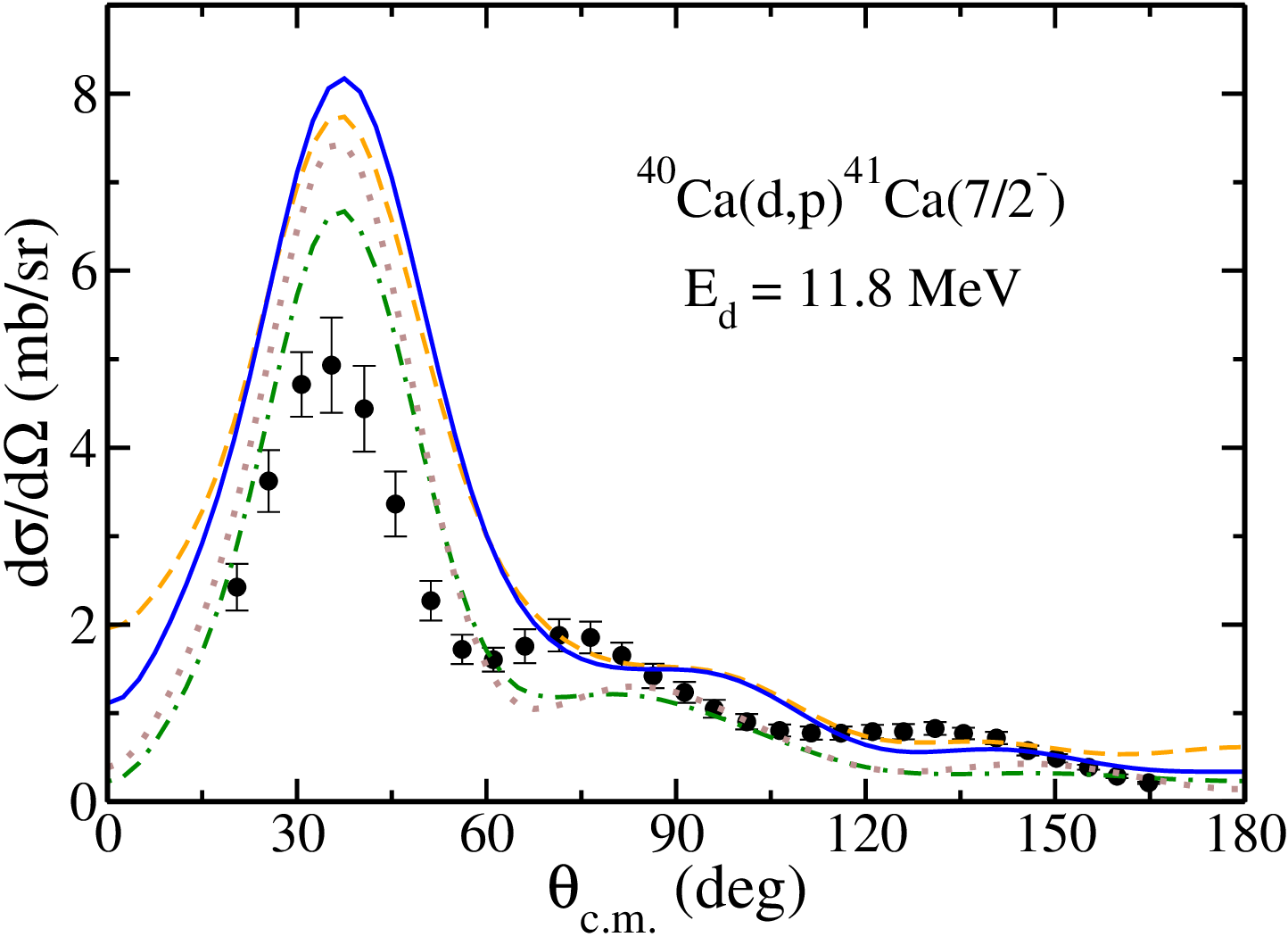}
\caption{(color online). Proton angular distributions for the $^{40}$Ca(d,p)$^{41}$Ca($7/2^-$) reaction with $E_d = 11.8$ MeV generated using the NLDOM (solid), GRZ (dashed), GR (dot-dashed), and TPM (dotted) nonlocal parameterizations.}
\label{Fig:xsec_nl4}
\end{figure}

\begin{center}
\begin{figure*}[t]
\includegraphics[width=0.9\textwidth]{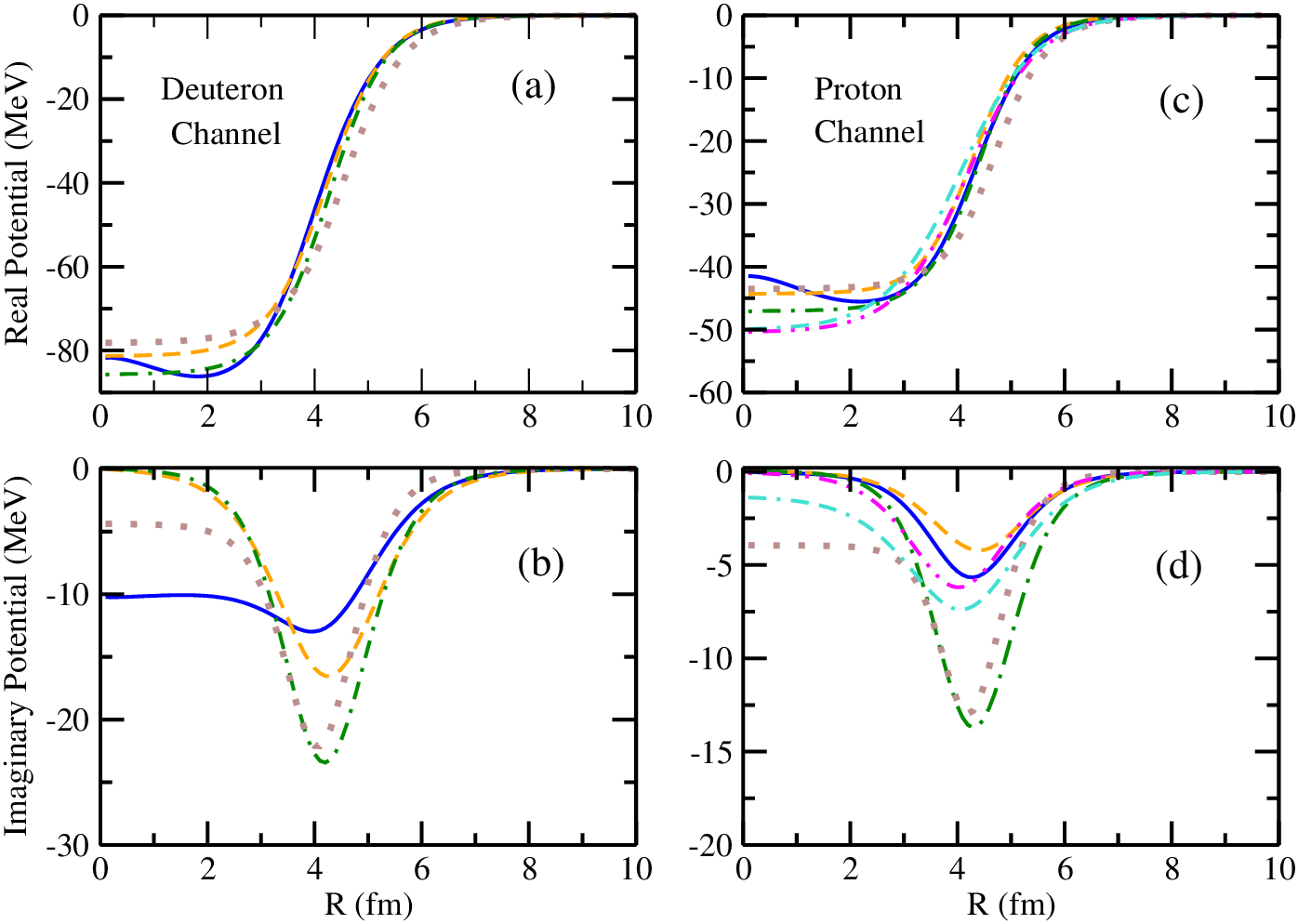}
\caption{(color online). Panels (a) and (b) show the real and imaginary parts of the entrance channel potentials for the NLDOM (solid), GRZ (dashed), GR (dot-dashed), and TPM (dotted) nonlocal parameterizations used in Fig. ~\ref{Fig:xsec_nl4}, while panels c) and d) show these quantities for the exit channel. Panels (c) and (d) also show the potentials calculated using LDOM (dot-dot-dashed) and CH89 (dash-dash-dotted). }
\label{Fig:pots_nl4}
\end{figure*}
\end{center}

Proton angular distributions calculated using the NLDOM potentials are presented in Fig.~\ref{Fig:xsec_dom_mod}.
The solid curve corresponds to the lowest-order result. The long-dashed curve shows that incorporating the first-order corrections for the proton channel, via Eqs. (\ref{Delta_U}) and (\ref{Perey_2}), reduces the lowest-order (d,p) peak cross sections $\sigma_{d,p}^{peak}$ by 3 \%. Further   first-order corrections, coming from Eqs. (\ref{delU}) and (\ref{f}) for the deuteron channel and shown by the short-dashed curve, reduces $\sigma_{d,p}^{peak}$ by another 5 \%.

Finally, including the spin-orbit potential raises $\sigma_{d,p}^{peak}$ and makes it comparable to the cross sections obtained with no corrections in the deuteron channel (the dot-dashed curve). The experimental data are from \cite{ca40}. The spread between all these calculations does not exceed 12\% and all of them considerably overestimate the experimental data, shown in Fig.~\ref{Fig:xsec_dom_mod} as well. This overestimation (by about 70\% after normalizing overlap function to 0.73) cannot come from the local approximations we have made to solve the nonlocal problem. The first-order corrections of about 12\% mean that the second-order corrections would most likely be around 1\% or less. Thus, other reasons for this overestimation should be investigated.

\begin{figure}[t]
\includegraphics*[width=0.4\textwidth]{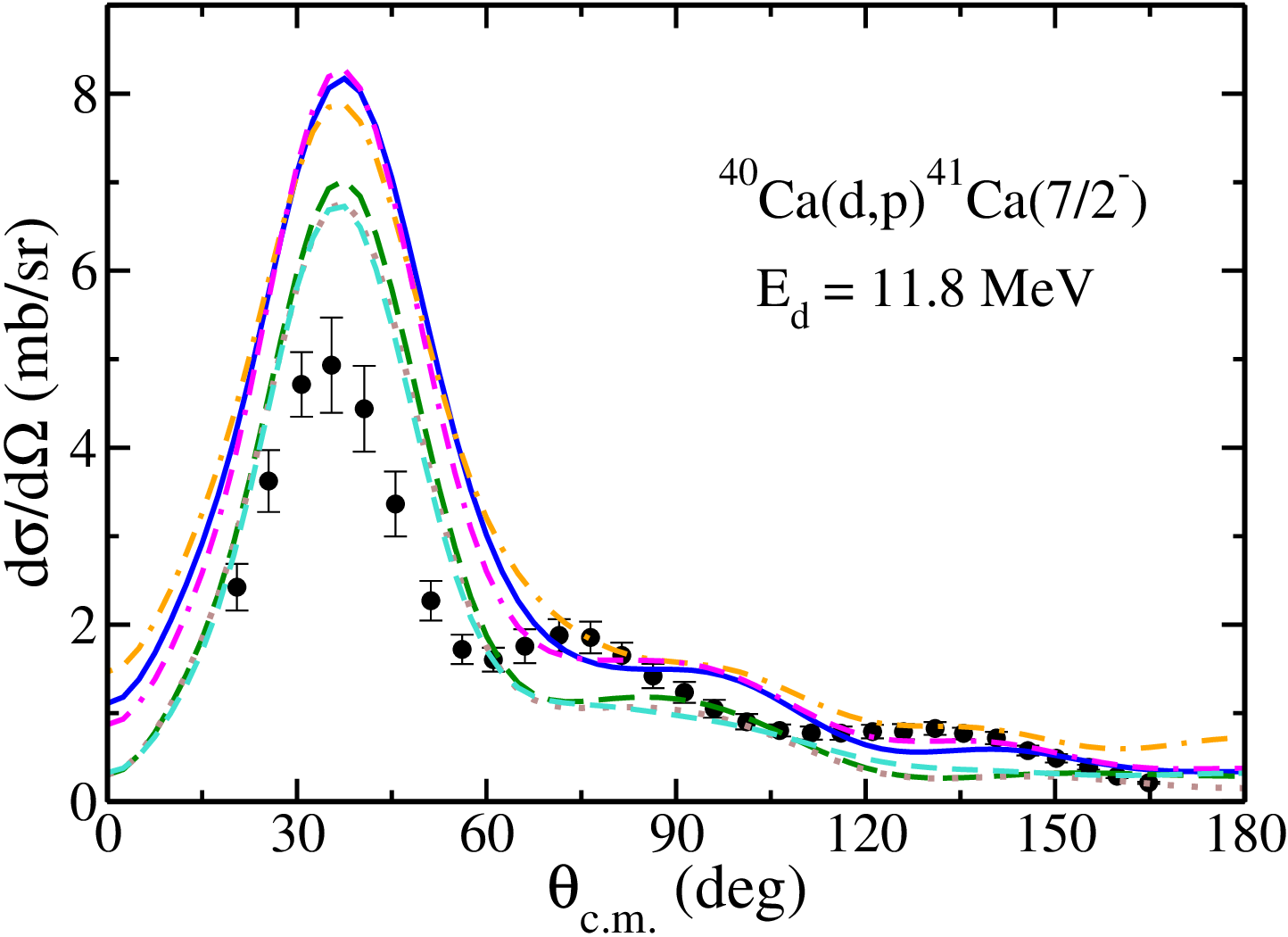}
\caption{(color online). Proton angular distributions for the $^{40}$Ca(d,p)$^{41}$Ca($7/2^-$) reaction with $E_d = 11.8$ MeV generated using NLDOM for the entrance channel and the overlap function and using NLDOM (solid), GRZ (dot-dashed), GR (long-dashed), TPM (dotted), LDOM (dash-dash-dotted), and CH89 (short-dashed) for the exit channel. }
\label{Fig:xsec_outps}
\end{figure}

It was already noted in \cite{Tim13a,Tim13b} that the adiabatic model with nonlocal energy-independent potentials gives higher cross sections, as compared to the standard adiabatic model, due to a weaker attraction in the deuteron channel. The higher cross sections are confirmed in other (d,p) studies with such potentials \cite{Titus16}. Here, the overestimation of the (d,p) cross sections using the NLDOM is stronger than in the case of energy-independent potentials. This can be seen in Fig.~\ref{Fig:xsec_nl4}, which compares the NLDOM angular distribution with the angular distributions from two nonlocal, energy-independent potentials, referred to as GR \cite{Gia76} and TPM \cite{TPM}. Figure~\ref{Fig:xsec_nl4} also shows the angular distribution from another nonlocal, energy-dependent potential, referred to as GRZ \cite{GRZ} and used previously in \cite{Joh14}. This potential generates an angular distribution very similar to the one generated with NLDOM. The structure of the GRZ potential is not as complicated as the NLDOM potential, but it has a typical low-energy behaviour of the imaginary part, vanishing at $E \rightarrow 0$ (for N = Z).  For all distributions, the corresponding nucleon optical model parameterizations were consistently used to calculate the exit and entrance channel potentials, taking into account the first-order corrections and the Perey effect. 

To understand the $\sim20\%$ difference in $\sigma_{d,p}^{peak}$ shown in Fig.~\ref{Fig:xsec_nl4}, we compare the entrance and exit channel potentials generated from the four nonlocal parameterizations and present them in 
Fig.~\ref{Fig:pots_nl4}. The NLDOM, GRZ and GR generate  real parts of similar depths and sizes both in entrance and exit channels while the  TPM produces a real part of  a moderately larger radial extent. 
The imaginary parts, however, show a marked difference, in both the entrance and exit channels. The energy-independent parameterizations GR and TPM produce a much larger imaginary part in the surface region than the energy-dependent parameterizations NLDOM and GRZ. The smaller imaginary parts  produce less  absorption thus increasing $\sigma_{d,p}^{peak}$. This is even better seen in
 Fig.~\ref{Fig:xsec_outps}, which shows the (d,p) angular distributions calculated with four
  different parameterizations for the exit proton channel and using
NLDOM for the deuteron channel. In this figure, we have also added the calculations with LDOM  \cite{Mueller11} and the widely used local CH89 \cite{CH89} parameterizations in the proton channel while keeping the NLDOM in the deuteron channel. The LDOM and CH89 proton potentials are shown in panels c) and d) of Fig.~\ref{Fig:pots_nl4}.

Figure~\ref{Fig:xsec_outps} shows that predictions with NLDOM, LDOM and GRZ form a different class from those obtained with GR, TPM and CH89. All the potentials from the first class have smaller imaginary parts and/or volume integrals than the potentials from the second class. Thus, overestimation of the cross section calculated with NLDOM seems to be at least partly due to a weaker absorption in the exit channel potential.
 
 \begin{figure}
\includegraphics*[width=0.4\textwidth]{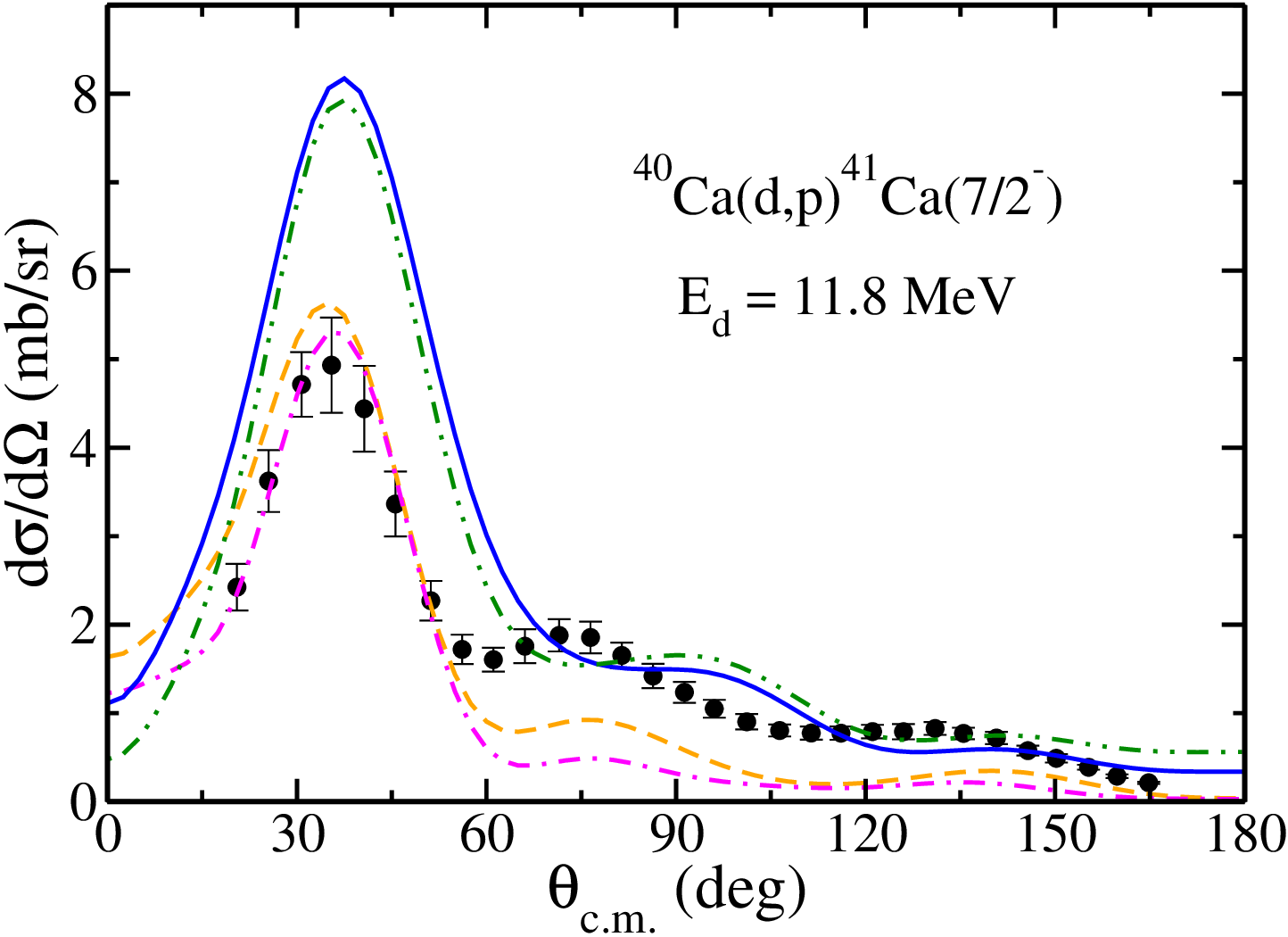}
\caption{(color online). Proton angular distributions for the $^{40}$Ca(d,p)$^{41}$Ca($7/2^-$) reaction at $E_d = 11.8$ MeV calculated with NLDOM (solid) and LDOM (dot-dot-dashed) using the TJ prescription, which evaluates the optical potentials at the shifted energy $E = E_d / 2 + 57$ MeV. Also shown are the results from using the JS prescription, which evaluates the optical potentials at the usual energy $E = E_d / 2$, with both NLDOM (dashed) and LDOM (dot-dashed). }
\label{Fig:js_tj_xsec}
\end{figure}

\begin{figure}[t!]
\includegraphics*[width=0.4\textwidth]{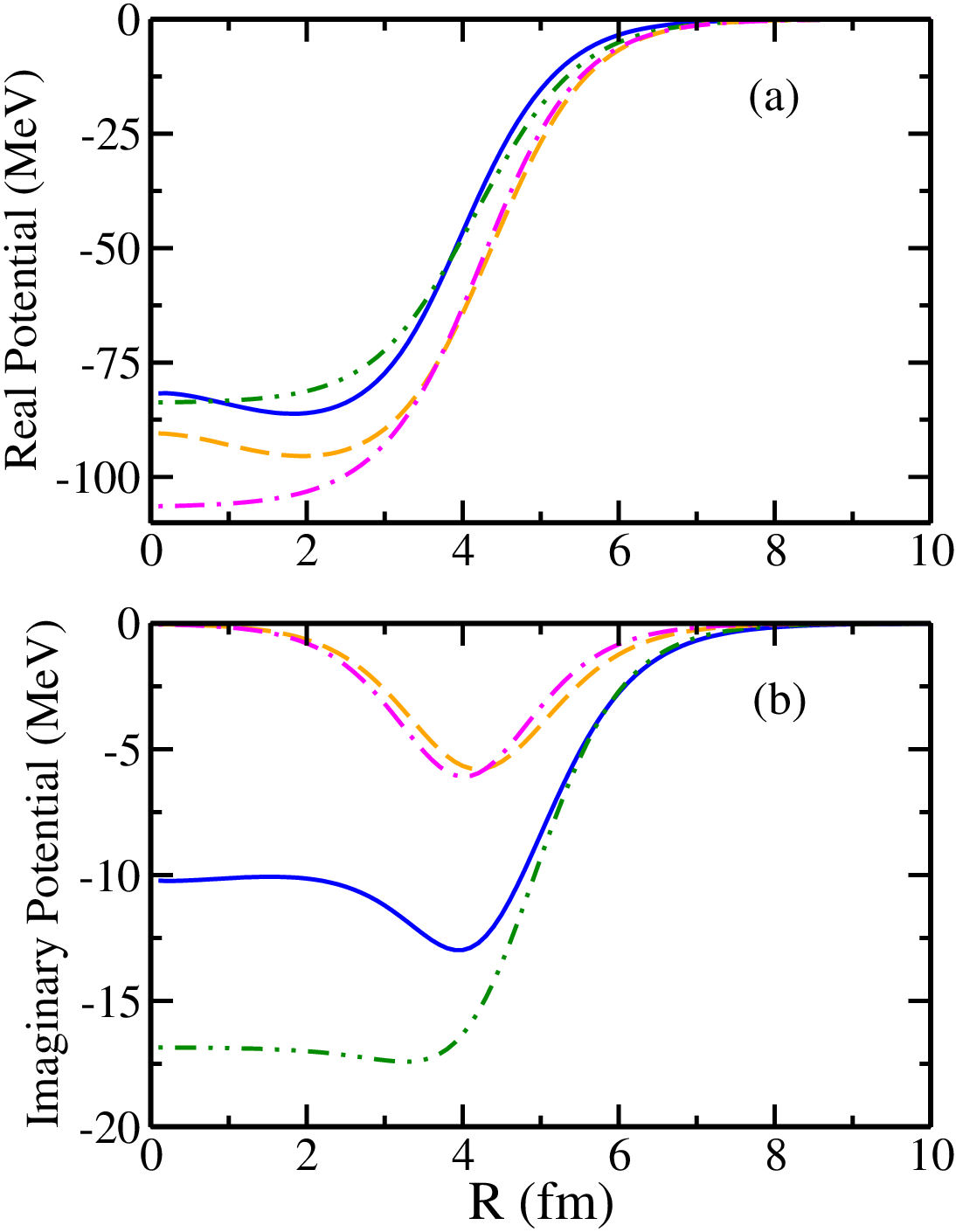}
\caption{(color online). (a) Real and (b) imaginary parts of the deuteron potential for the $^{40}$Ca(d,p)$^{41}$Ca($7/2^-$) reaction at $E_d = 11.8$ MeV calculated with NLDOM (solid) and LDOM (dot-dot-dashed) using the TJ prescription, which evaluates the optical potentials at the shifted energy $E = E_d / 2 + 57$ MeV. Also shown are the results from using the JS prescription, which evaluates the optical potentials at the usual energy $E = E_d / 2$, with both NLDOM (dashed) and LDOM (dot-dashed).}
\label{Fig:js_tj_pots}
\end{figure}
 
The weaker absorption may not be the only reason for large cross sections obtained with NLDOM. It was discussed in detail in \cite{Har71} that a particular relation between optical potentials in entrance and exit channels results in destructive interference between the ingoing and outgoing partial waves leading to $l$-localization of radial (d,p) amplitudes in the adiabatic model. A similar situation may occur here. Indeed, the standard adiabatic Johnson-Soper (JS) \cite{JS} calculations using both the local-equivalent NLDOM and LDOM potentials, taken at $E=E_d/2$, predict much lower cross sections (see Fig.~\ref{Fig:js_tj_xsec}) while the imaginary parts of the JS potentials, shown in Fig.~\ref{Fig:js_tj_pots}, are much smaller than those of local-equivalent deuteron potentials obtained in this work (TJ). This could be an indication of constructive interference between the ingoing and outgoing partial waves generated with NLDOM potentials. A new procedure was proposed in Sec. VI.B of Ref.~ \cite{Joh14}, explaining how phenomenological local energy-dependent optical potentials can be used in (d,p) calculations if they represent local equivalents of nonlocal potentials. Using the LDOM potential within this procedure and assuming a hidden nonlocality of 0.85 fm gives very similar results to NLDOM both for the real part of deuteron distorting potential (Fig.~\ref{Fig:js_tj_pots}) and the (d,p) cross sections (Fig.~\ref{Fig:js_tj_xsec}) despite stronger imaginary part in the deuteron channel. Although the JS cross sections are close to experimental data, in light of recent findings \cite{Tim13a,Tim13b, Joh14,Titus16}, constructing the adiabatic potentials from nucleon optical potentials taken at half the deuteron incident energy does not seem to be justified anymore.  

\section{Conclusion}
\label{sec:con}

We presented the first adiabatic (d,p) calculations with the NLDOM potential, which has been designed with the aim of forging the link between nuclear structure and nuclear reactions in a consistent way. It has its roots in the underlying self-consistent Green's functions theory and possesses the fundamental properties - nonlocality, energy-dependence and dispersion relations -  that arise from the complex structure of the target. The NLDOM explicitly takes into account a number of components of nuclear many-body theory that many other optical models do not. 

One could expect that using an advanced optical potential parametrization such as NLDOM would result
in properly fixed single-nucleon properties both below and above the Fermi surface crucial for
agreement between predictions of (d,p) reaction theory and experimental data. However, we have shown that using the NLDOM to generate the distorting potentials entering the (d,p) amplitude strongly overestimates the (d,p) cross sections despite the reduced strength of the NLDOM one-neutron overlap function employed in the calculations. Moreover, the NLDOM predictions are very similar to those made with a much simpler nonlocal potential GRZ derived within Watson multiple scattering theory and Wolfenstein's parameterization of the nucleon-nucleon scattering amplitude \cite{Gia76,GRZ}. The energy dependence is presented in GRZ only in the imaginary part.

Since we do not have strong reason to doubt the quality of the NLDOM parameterization the main assumptions of the (d,p) theory used in the present calculations should be reviewed. We list them below:
\begin{itemize}
 \item The (d,p) amplitude contains a projection of the total many-body wave function into the three-body channel $A+n+p$ only. Projections onto all excited states of A are neglected.
 \item Only $n-A$ and $p-A$ potentials are used to calculate the $A+n+p$ projection. According to \cite{Joh14} there are also multiple scattering terms playing the role of a three-body $A+n+p$ force. These are neglected.
 \item Averaged $n-A$ and $p-A$ potentials were obtained using the procedure from Ref. \cite{Joh14}, which uses the adiabatic approximation. Corrections to this approximation may change the energy value at which these potentials should be evaluated.
 \item It was assumed that the (d,p) transition operator contains the $V_{np}$ term only. Any other terms present in this amplitude \cite{Satchler} are neglected.
 \item It was shown in \cite{Tim99} that keeping $V_{np}$ only in the (d,p) transition operator modifies the proton channel wave function. In our particular case, this would result in using  p-$^{40}$Ca optical potential in the p+$^{41}$Ca channel. We have not seen any difference in (d,p) cross sections when replacing $^{41}$Ca by $^{40}$Ca and this could mean that the averaging procedure, introduced in \cite{Joh14}, when applied to
 the special $A+n+p$ three-body model, that does not have $V_{np}$ and has different asymptotic conditions \cite{Tim99}, may result in completely different requirements to the proton distorting potential in the exit channel. Using proton optical potentials may not be justified anymore.
 \item We used the adiabatic approximation to solve the three-body Schr\"odinger equation.  
\end{itemize}
The deviation from the adiabatic approximation for solving the Schr\"odinger equation has been studied both within the continuum-discretized coupled channel method \cite{Mor09} and using Faddeev equations \cite{Nunes11}. Although these corrections can be non-negligible, they cannot be responsible for 70\% overestimation of (d,p) cross sections obtained in this work. At $E_d/2 \sim 12$ MeV these deviations were no more than 4$\%$, while at a larger energy range, $5\leq E\leq 56$ MeV they could be up to 23$\%$. The unknown non-adiabatic corrections to optical $n-A$ and $p-A$ potentials entering the Schr\"odinger equation for the $A+n+p$ model  \cite{Joh14} can change both real and imaginary parts of these effective potentials, which could affect the (d,p) cross sections. But given that the adiabatic approximation is a good first choice for the (d,p) reactions, most likely, they will not explain the 70\% difference between the NLDOM predictions and experiment.

The contributions from the remnant term in the (d,p) amplitude (all other terms that are not $V_{np}$) have been studied in an inert core model, where they were found to be small \cite{Smith69}. We estimated the effect of the remnant term for the $^{40}$Ca(d,p)$^{41}$Ca reaction at $E_d = 11.8$ MeV using FRESCO \cite{FRESCO}, which employs the inert core approximation and requires the use of local-equivalent potentials. We also found the effect of the remnant term to be small, decreasing the cross-section by about 3$\%$. A more recent study \cite{Ramos15} showed that the remnant term contributions remain small even when incorporating core-excitation effects, although they can become more important for nuclei in which the core has a low excitation energy. Whether these contributions remain small for a nonlocal $U_p$ is not known.

The strong overestimation of the $^{40}$Ca(d,p)$^{41}$Ca cross sections at 11.8 MeV implies that neglected parts of the (d,p) amplitude and/or its constituents are much more important than was thought before. Given that the deuteron energy, chosen for this work, is often used in modern experiments with radioactive beams for spectroscopic and astrophysical reasons, and that the dispersive effects are strong at this energy, further development of direct reaction theories is crucial to understand transfer experiments performed either recently or in the past and planned for the future.

\section*{Acknowledgement}

N.K.T.  gratefully acknowledges support from the UK STFC through Grant ST/L005743/1. We are grateful to R.C. Johnson for fruitful discussions. S. J. Waldecker would also like to thank W. H. Dickhoff and M. H. Mahzoon for fruitful discussions.

\end{document}